%% file: ms.tex
\documentclass[12pt,preprint]{aastex}

\shorttitle{ The Evolution of the Ly$\alpha$ Photon Escape Fraction}
\shortauthors{Blanc et al.}

\begin{document}

\title{The HETDEX Pilot Survey. II. The Evolution of the L\lowercase{y}$\alpha$ Escape Fraction from 
the UV Slope and Luminosity Function of $1.9<$\lowercase{z}$<3.8$ LAE\lowercase{s}}

\author{Guillermo A. Blanc,\altaffilmark{1} Joshua J.
  Adams,\altaffilmark{1} Karl Gebhardt,\altaffilmark{1,2} Gary J.
  Hill,\altaffilmark{2,3} Niv Drory\altaffilmark{4}, Lei Hao\altaffilmark{1,5},
Ralf Bender\altaffilmark{4,6},
Robin Ciardullo\altaffilmark{7},
Steven L. Finkelstein\altaffilmark{8}, Alexander B. Fry\altaffilmark{1,13},
Eric Gawiser\altaffilmark{9}, Caryl Gronwall\altaffilmark{7},
Ulrich Hopp\altaffilmark{4,6}, Donghui Jeong\altaffilmark{1,2,10},
Ralf Kelzenberg\altaffilmark{4}, Eiichiro Komatsu\altaffilmark{1,2},
Phillip MacQueen\altaffilmark{3},
Jeremy D. Murphy\altaffilmark{1},
Martin M. Roth\altaffilmark{11}, Donald P. Schneider\altaffilmark{7},
Joseph Tufts\altaffilmark{3,12}}

\altaffiltext{1}{Department of Astronomy, University of Texas at Austin, Austin, TX}
\altaffiltext{2}{Texas Cosmology Center, University of Texas at Austin, Austin, TX}
\altaffiltext{3}{McDonald Observatory, Austin, TX}
\altaffiltext{4}{Max Planck Institute for Extraterrestrial Physics, Garching, Germany}
\altaffiltext{5}{Current Address: Shanghai Astronomical Observatory, Shanghai, China}
\altaffiltext{6}{University Observatory Munich, Munich, Germany}
\altaffiltext{7}{Department of Astronomy and Astrophysics,
  Pennsylvania State University, University Park, PA}
\altaffiltext{8}{George P. and Cynthia Woods Mitchell Institute for
  Fundamental Physics and Astronomy, Department of Physics and
  Astronomy, Texas A\&M University, College Station, TX}
\altaffiltext{9}{Department of Physics and Astronomy, Rutgers University, Piscataway, NJ}
\altaffiltext{10}{Current Address: California Institute of Technology, Pasadena, CA}
\altaffiltext{11}{Astrophysikalisches Institut Potsdam, Postdam, Germany}
\altaffiltext{12}{Current Address: Las Cumbres Observatory, Santa Barbara, CA}
\altaffiltext{13}{Current Address: The University of Washington, Seattle, WA}

\begin{abstract}

We study the escape of Ly$\alpha$ photons from Ly$\alpha$ emitting
galaxies (LAEs) and the overall galaxy population using a
sample of 99 LAEs at $1.9<z<3.8$ detected through integral-field
spectroscopy of blank fields by the HETDEX Pilot Survey. For 89 LAEs
with broad-band counterparts we measure UV luminosities and UV slopes,
and estimate $E(B-V)$ under the assumption of a constant intrinsic UV
slope for LAEs. These quantities are used to estimate dust-corrected
star formation rates ($SFR$). Comparison between the observed
Ly$\alpha$ luminosity and that predicted by the dust-corrected $SFR$
yields the Ly$\alpha$ escape fraction. We also measure the Ly$\alpha$
luminosity function and luminosity density ($\rho_{Ly\alpha}$) at
$2<z<4$. Using this and other measurements from the literature at
$0.3<z<7.7$ we trace the redshift evolution of $\rho_{Ly\alpha}$. We
compare it to the expectations from the star-formation history of the
universe and characterize the evolution of the Ly$\alpha$ escape
fraction of galaxies. LAEs at $2<z<4$ selected
down to a luminosity limit of $L(Ly\alpha)>3-6\times10^{42}$
erg s$^{-1}$ (0.25 to 0.5 $L^*$), have a mean
$\langle E(B-V) \rangle=0.13\pm0.01$, implying an attenuation of
$\sim70$\% in the UV. They show a median UV uncorrected $SFR=
11$ M$_{\odot}$yr$^{-1}$,  dust-corrected $SFR=34$
M$_{\odot}$yr$^{-1}$, and Ly$\alpha$ equivalent widths ($EW$s) which are
consistent with normal stellar populations. 
We measure a median Ly$\alpha$ escape fraction of $29$\%, with a large
scatter and values ranging from a few percent to 100\%. 
The Ly$\alpha$ escape fraction in LAEs correlates with
$E(B-V)$ in a way that is expected if Ly$\alpha$
photons suffer from similar amounts of dust extinction
as UV continuum photons. This result implies that a strong enhancement of the
Ly$\alpha$ $EW$ with dust, due to a clumpy multi-phase ISM,
is not a common process in LAEs at these redshifts. It also suggests
that while in other galaxies Ly$\alpha$ can be preferentially quenched
by dust due to its scattering nature, this is not the case in LAEs.
We find no evolution in the average dust content and Ly$\alpha$ escape
fraction of LAEs from $z\sim 4$ to 2. We see hints of a drop in the
number density of LAEs from $z\sim4$ to 2 in the redshift distribution
and the Ly$\alpha$ luminosity function, although larger samples are
required to confirm this. The mean Ly$\alpha$ escape
fraction of the overall galaxy population decreases significantly from
$z\sim6$ to $z\sim2$. Our results point towards a scenario in which
star-forming galaxies build up significant amounts of dust in
their ISM between $z\sim6$ and 2, reducing their Ly$\alpha$ escape
fraction, with LAE selection preferentially detecting galaxies
which have the highest escape fractions given their dust content. The fact
that a large escape of Ly$\alpha$ photons is reached by $z\sim6$
implies that better constraints on this quantity at higher redshifts
might detect re-ionization in a way that is uncoupled from
the effects of dust.

\email{gblancm@astro.as.utexas.edu}

\end{abstract}


\section{Introduction}

Ly$\alpha$ photons are produced in large amounts in star forming
regions, therefore it was predicted nearly half a century ago that
the Ly$\alpha$ emission line at 1216\AA\ should be a signpost for
star-forming galaxies at high redshift \citep{partridge67}. Actual
observations of Ly$\alpha$ emitting (LAE) galaxies at high redshift had to
wait for the advent of 8-10m class telescopes \citep{hu98}. A
little more than a decade has passed since their discovery, and thanks to a
series of systematic surveys at optical and near-infrared wavelengths,
large samples of LAEs, usually containing from tens to a few hundred
objects, have been compiled over a wide range of redshifts from
$z\sim2$ to $z\sim7$ \citep[eg.][]{cowie98, rhoads00, kudritzki00, malhotra02, ouchi03, gawiser06a,
ajiki06, gronwall07, ouchi08, nilsson09a, finkelstein09, guaita10,
hayes10a, ono10, adams11}. Space based ultra-violet (UV) observations have also
been used to study Ly$\alpha$ emitting galaxies at lower redshifts, all
the way down to the local universe \citep{kunth98, kunth03, hayes05,
  hayes07, atek08, deharveng08, cowie10}.

The intrinsic production of both Ly$\alpha$ and UV continuum photons
in a galaxy is directly proportional to the
number of ionizing photons produced by young stars,
which is proportional to the star formation rate ($SFR$) \citep{kennicutt98,
  schaerer03}. In practice, we do
not expect the observed Ly$\alpha$ luminosity of galaxies to
correlate well with their $SFR$ because the resonant nature of
the n=1-0 transition in hydrogen makes the escape of Ly$\alpha$
photons a non-trivial radiative process. 

In principle, the large number of
scatterings suffered by a Ly$\alpha$ photon before escaping the
neutral medium of a galaxy increase its probability, with respect to
that of continuum photons outside the resonance wavelength, of being absorbed
by a dust grain. Hence, we would expect even small amounts of dust in
a galaxy's inter-stellar medium (ISM) to severely decrease the
equivalent width ($EW$) of the Ly$\alpha$ line \citep{hummer80, charlot93}. In
reality the situation is far more complicated, and it is not clear how
the extinction
suffered by Ly$\alpha$, and that suffered by continuum photons,
relate. One
scenario which has been proposed by several authors
\citep{neufeld91, haiman99, hansen06} is the possible enhancement of
the Ly$\alpha$ $EW$ due to the presence of a very clumpy dust
distribution in a multi-phase
ISM. For this type of ISM geometry most of the dust lives in cold
neutral clouds embedded in an ionized medium. In this scenario,
Ly$\alpha$ photons have a high probability of being scattered in the
surfaces of these clouds, spending most of their time prior to escape in the
inter-cloud medium and actually suffering less dust extinction than
non-resonant radiation, which can penetrate into the clouds where it
has a higher chance of being absorbed or scattered by dust
grains. Recently, \cite{finkelstein09} claimed that this
process can simultaneously explain the Ly$\alpha$ fluxes and
continuum spectral energy distributions of many objects in their sample of
LAEs at $z \sim 4.5$.

At high redshift the Ly$\alpha$ line can also be affected by scattering in
the inter-galactic medium (IGM), as escaping Ly$\alpha$ photons
bluewards of the line center can be redshifted into the resonance
wavelength. This effect is particularly important at $z > 5$ as the
density of neutral gas in the universe
increases, but even at lower redshifts, when the universe is almost
completely ionized, intervening Ly$\alpha$ forest absorption can occur. 
To first order, the IGM transmission blue-wards of Ly$\alpha$
is $\sim 90$\%, 70\%, and 50\% at $z\sim1.9$, 3.0, and 3.8
respectively \citep{madau95}. In
the naive case where the line profile escaping a galaxy is symmetric
and centered at the Ly$\alpha$ resonance, since only photons bluewards
of the line are affected, we can expect attenuations of $\sim 5$\%,
15\%, and 25\% on the emerging flux at these redshifts. In reality the
process can be
significantly different. While inflow of IGM gas onto galaxies can
introduce further attenuation red-wards of the line resonance
\citep{dijkstra07}, outflows in a galaxy's ISM can redshift the
emerging spectrum so as to be completely unaffected by the IGM
\citep{verhamme08}. For example, in a sample of 11 LBGs and LAEs at
$z\sim 3 - 5$, \cite{verhamme08} find no need to introduce IGM
absorption to successfully fit the observed line profiles. This,
combined with the inherent stochasticity of intervening absorption
systems towards different lines of sight, makes Ly$\alpha$ IGM attenuation
corrections very difficult and uncertain.

The kinematics of the neutral gas inside a galaxy and in its immediate
surroundings also play an important role regarding the escape of
Ly$\alpha$ photons \citep{verhamme06, dijkstra06, hansen06, dijkstra07,
  verhamme08, adams09, laursen10, zheng10}. Simply put, the velocity field of the neutral gas has a
strong influence on the emission line profile of the Ly$\alpha$
line. Different combinations of geometry and velocity fields can ``move''
photons out of the resonance frequency either by blueshifting
(typically due to in-falling gas) or
redshifting (due to outflows) them, changing the number of scatterings photons
experience before exiting the galaxy as well as their escape frequency. This
process can affect the amount of dust extinction as well as
the amount of potential IGM scattering those photons will suffer.

No clear agreement is found in the literature regarding the amount of
dust present in the ISM of Ly$\alpha$ emitting galaxies. While most studies of
narrow-band selected LAEs at $z\sim3$ seem to indicate they are
consistent with very low dust or dust-free stellar populations
\citep{gawiser06a, gawiser07, nilsson07, gronwall07, ouchi08}, there have been
recent results suggesting that the LAE population is more
heterogeneous and includes more dusty and evolved galaxies,
especially at lower redshifts \citep{lai08, nilsson09a, finkelstein09}.

We use a new sample of spectroscopically detected LAEs at
$1.9<z<3.8$ from the The Hobby Eberly Telescope Dark Energy Experiment
(HETDEX) Pilot Survey \citep{adams11} to investigate the shape of the UV continuum of LAEs,
as well as the Ly$\alpha$ luminosity function of these objects, and to
address:

\begin{itemize}
\item The dust content of LAEs, parameterized by the dust reddening $E(B-V)$,
  and its evolution with redshift.
\item The star-formation properties ($SFR$), the Ly$\alpha$ escape
  fraction in LAEs, and its evolution with redshift.
\item The relation between the dust content and the escape fraction of
Ly$\alpha$ photons.
\item The relation between the dust extinction seen by continuum and
  resonant Ly$\alpha$ photons.
\item The contribution of LAEs to the integrated star formation rate
  density at different redshifts.
\item The Ly$\alpha$ escape fraction of the overall galaxy
  population and its evolution with redshift.
\end{itemize} 

These galaxies have been detected through wide integral-field
spectroscopic mapping of blank fields, using the Visible
Integral-field Replicable Unit Spectrograph Prototype
\citep[VIRUS-P,][]{hill08}. The Pilot Survey catalog of emission line
galaxies is presented in \cite{adams11}, hereafter Paper I.
The large redshift range spanned by
our sample allows us to check for any potential evolution in the
above properties of LAEs.

In \S2 we describe the HETDEX Pilot Survey from which the sample of
Ly$\alpha$ emitting galaxies is drawn. In \S3 we present
our sample of LAEs along with their luminosities and
redshift distribution. \S4 presents our measurement of the
UV continuum slope and derivation of the amount of dust extinction
present in these objects. Discussion of any potential evolution in the
dust properties of LAEs is also in this section. We compare both
uncorrected as well as dust-corrected $SFR$s derived
from both UV and Ly$\alpha$ in \S5, where we also compute the escape
fraction of Ly$\alpha$ photons and show how it depends on the amount of
dust reddening. In \S6 we present the Ly$\alpha$ luminosity function and 
check for its possible evolution with redshift. We compare the
integrated $SFR$ density derived from the Ly$\alpha$ luminosity function
to that for the global galaxy population in \S7. In this way we can assess
the contribution of LAEs to the star-formation budget of the universe
at these redshifts and estimate the Ly$\alpha$ photon escape fraction
for the overall galaxy population. Finally, we
summarize our results and present our conclusions in \S8.

Throughout the paper we adopt a standard set of $\Lambda$CDM cosmological
parameters, $H_o=70$ km s$^{-1}$Mpc$^{-1}$, $\Omega_M=0.3$,
and $\Omega_{\Lambda}=0.7$ \citep{dunkley09}.

\section{The HETDEX Pilot Survey}

Ever since their discovery, the standard method for detecting and
selecting LAEs has been through narrow-band imaging in a passband
sampling the Ly$\alpha$ line at a given redshift. The redshift range of these
type of surveys is given by the width of the narrow-band filter used, and
is typically of the order of $\Delta z=0.1$. Hence, these studies are
limited to very narrow and specific redshift ranges. In terms of
surveyed volumes this limitation is compensated by the large fields of view of
currently available optical imagers which allow for large areas of the
sky ($\sim 1 $deg$^2$) to be surveyed using this technique.

An alternative technique, which has been attempted for detecting LAEs over
the last few years, is to do so through blind spectroscopy. This can be
done either by performing very low resolution slit-less spectroscopy
\citep{kurk04, deharveng08}, blind slit spectroscopy
\citep{martin04, tran04, rauch08, sawicki08, cassata11}, or integral-field
spectroscopy \citep{vanbreukelen05}. 

The success of this type of surveys has been variable. While early
attempts to detect LAEs at $z\sim6$ using slit spectroscopy failed to
do so, and could only set upper limits to their number density \citep{martin04,
  tran04}, more recent attempts at lower redshifts ($2<z<6$) like the
ones by \cite{rauch08} and \cite{cassata11}, have
produced large samples of objects. Similarly, an early attempt by
\cite{kurk04} to find LAEs at $z=6.5$ using slit-less spectroscopy
only yielded one detection, while more recently space-based UV
slit-less spectroscopy with the GALEX telescope has allowed for the
construction of a large sample of LAEs at $z\sim0.3$ \citep{deharveng08}.
The only attempt to detect LAEs using integral-field spectroscopy
previous to this work was done by \cite{vanbreukelen05}, who
used the Visible Multi-Object Spectrograph (VIMOS) integral
field unit (IFU) on the Very Large Telescope (VLT) to build a sample
of 18 LAEs at $2.3<z<4.6$ over an area of 1.44 arcmin$^2$
corresponding to the VIMOS IFU field-of-view.

Although when doing spectroscopic searches for LAEs the wavelength
range, and hence the redshift over which Ly$\alpha$ can be detected,
is tens of times larger than for narrow-band imaging,
surveyed volumes have been typically small due to the small areas
sampled by the slits on the sky, or the small fields-of-view of most
integral field units. For example, the IFU survey by
\cite{vanbreukelen05}
only covered $\sim 10^4$Mpc$^3$ because of the small area
surveyed, while the $z\simeq3.1$ narrow-band survey by
\cite{gronwall07} covered $\sim 10^5$Mpc$^3$ over a very narrow range
of $\Delta z=0.04$ because of the large $36'\times36'$ area which
can be imaged with the MOSAIC-II camera. It is clear that the most
efficient way of building large samples of LAEs would be to conduct
spectroscopic searches over large areas of the sky.

HETDEX \citep{hill08} will survey $\sim60$ deg$^2$ of sky\footnote{The actual HETDEX footprint
  corresponds to a 420 deg$^2$ area, but only 1/7 of the field will be
covered by fibers} using the Visible Integral-field
Replicable Unit Spectrograph \citep[VIRUS,][]{hill10}, a wide field of view ($16'\times
16'$) integral field spectrograph currently being built for the 9.2m
Hobby Eberly Telescope (HET). HETDEX will produce a sample of
$\sim8\times 10^5$ LAEs at $1.9<z<3.5$ over a volume of
8.7 Gpc$^3$. The power-spectrum of the spatial
distribution of these objects will be used to set a percent level constrain
on the dark energy equation of state parameter $w$ at these high
redshifts \citep{hill08}. A prototype of the instrument, VIRUS-P,
is currently the largest field-of-view IFU in existence, and has been used
over the last 3 years to conduct a Pilot Survey for LAEs from which
the sample used in this work is taken from (Paper I). The Pilot Survey, described
below, samples the $1.9<z<3.8$ range, and covers a volume of
$\sim10^6$ Mpc$^3$ over an area of 169 arcmin$^2$. This volume is ten
times larger than the one covered in \cite{gronwall07} and
\cite{guaita10}, three times larger than the one covered by
\cite{nilsson09a}, and of comparable size to the one sampled at $z=3.1$
by \cite{ouchi08} but over an area 20 times smaller, exemplifying the
power of integral field spectroscopy to search for emission line
galaxies over large volumes.

The HETDEX Pilot Survey obtained integral field spectroscopy over $\sim$169.23
arcmin$^2$ of blank sky in four extra-galactic fields \citep[COSMOS: 71.6
arcmin$^2$, GOODS-N: 35.5 arcmin$^2$, MUNICS-S2: 49.9 arcmin$^2$, and
XMM-LSS: 12.3 arcmin$^2$;][]{scoville07,dickinson03,drory01,pierre04} using 
VIRUS-P on the 2.7m Harlan J. Smith telescope at
McDonald Observatory. The goal of the survey is to conduct an
unbiased search for spectroscopically-detected emission line galaxies
over a wide range of redshifts. Although a powerful dataset itself, the
Pilot Survey also provides a proof of concept and a crucial
test-bench for the planned HETDEX survey.

The observations and data reduction, as well as the detection and
classification of emission line galaxies, are presented in
Paper I, and we refer the reader to it for a more detailed description
of the survey design. Briefly, each field is mapped by a mosaic of
$1.7'\times 1.7'$ VIRUS-P pointings (27, 13, 16, and 4 pointings in
COSMOS, GOODS-N, MUNICS-S2, and XMM-LSS respectively). The VIRUS-P IFU
consists of an square array of 246 fibers, each 4.235$''$ in diameter,
sampling the field with a 1/3 filling factor. While
a set of three dithered exposures covers the field-of-view almost
completely, we observed each
pointing at six dithered positions, ensuring complete coverage
and improving the spatial sampling of the field and the astrometric accuracy of
our detections. For each pointing, we obtained spectra at 1,476
($6\times 246$) positions, with any point on the sky being typically
sampled by 2 overlapping fibers. Overall, the Pilot Survey
consists of $\sim$88,000 individual spectra over 169 arcmin$^2$ of blank
sky. Each spectrum covers the 3600\AA -5800\AA\ wavelength range with
$\sim 5$\AA\ FWHM resolution ($\sigma_{inst}\sim 130$ km s$^{-1}$ at 5000\AA).

After the data are reduced and a 1D flux-calibrated spectrum is extracted for each fiber
position, we search the ``blank'' spectra for emission lines using an
automated procedure (Paper I). Line detections are
associated, when possible, with counterparts in broad-band images
available for all four fields. The VIRUS-P wavelength range
allows the detection of common strong emission lines present in
star-forming galaxies such as Ly$\alpha$ at $1.9<z<3.8$, [OII]$\lambda$3727 at
$z<0.56$, H$\beta$ at $z<0.19$, [OIII]$\lambda$4959 at $z<0.17$,
[OIII]$\lambda$5007 at $z<0.16$, as well as typical AGN lines like
CIV$\lambda$1549 at $1.3<z<2.7$, CIII]$\lambda$1909 at $0.9<z<2.0$,
and MgII$\lambda$2798 at $0.3<z<1.1$.

Source classification is based on the presence of multiple spectral
lines when available. In the case of single line detections, the spectral
classification is considerably more challenging. For LAEs, only the Ly$\alpha$ line
appears in our wavelength range, so we expect single line
detections for our objects of interest. Nevertheless, [OII] emitters at
$0.19<z<0.56$ will also appear as single line detections in the
VIRUS-P spectra. Even [OII] emitters at $z<0.19$ that have unfavorable
emission line ratios can appear as single line detections if H$\beta$ and
the [OIII] doublet are below the noise level. Our 5\AA\ FWHM spectral resolution is not high enough
to resolve the [OII]$\lambda$3727 doublet, so we cannot rely on the
line profile to classify these objects. While galaxies detected in
redder lines such as H$\beta$ and [OIII]$\lambda$5007 can also appear as
single line detections depending on their redshifts and line ratios,
the volume over which we sample these galaxies is $\sim400$ times
smaller than the volume over which we sample LAEs, and $\sim20$ times
smaller than the volume over which we sample [OII] emitters. Hence,
contamination from H$\beta$ and [OIII] emitters is negligible.

The classification of single line detections is thoroughly discussed
in Paper I, and is based on an $EW$ criterion, where
objects showing rest-frame $EW(Ly\alpha)>20$\AA\ are classified as LAEs
(for 4 objects the $EW>20$\AA\ criterion was bypassed due to the existence of
further evidence pointing towards their LAE nature; see Paper I).
This $EW$ constraint effectively reduces the contamination from low-z interlopers to a
negligible level. A total of 105 Ly$\alpha$ detections are present in
the Pilot Survey catalog presented in Paper I. Of these, 6 show X-ray
counterparts indicating an AGN nature, leaving a final sample of 99
``normal'' star-forming LAEs. In Paper I we also present a thorough
assessment of the completeness and spurious source contamination in
our catalog, based on simulated data. The completeness is used in \S6
to estimate the Ly$\alpha$ luminosity function. In our sample of LAEs we expect a
4-10\% contamination from spurious sources. The sample used in this
work is presented in Table \ref{tbl-1}.

\section{LAE Sample}

\begin{figure}[t]
\begin{center}
\plotone{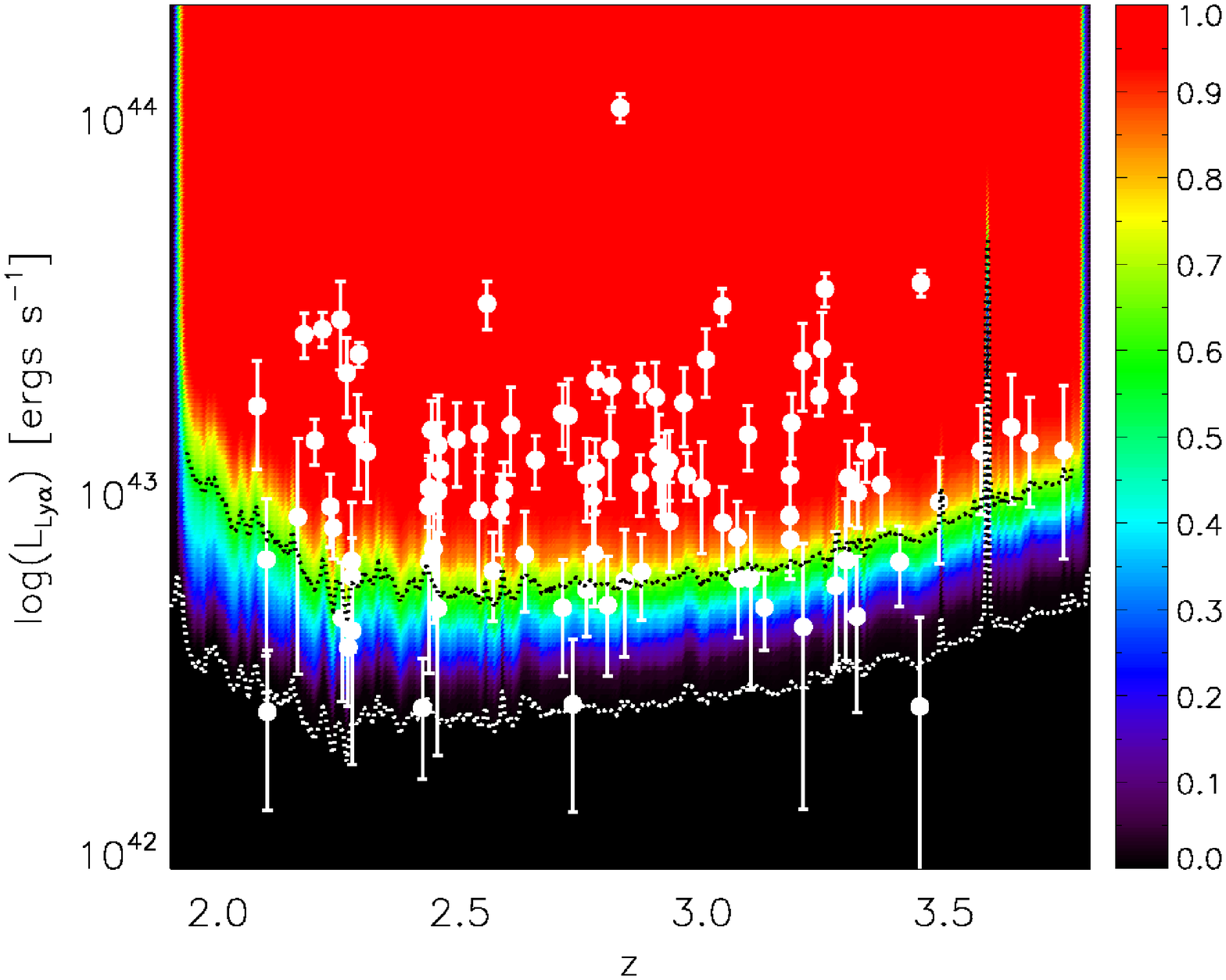}
\caption{Limiting Ly$\alpha$ luminosity ($5\sigma$) as a function of
  redshift for the survey. The survey depth varies across the observed
  area due to changes in atmospheric transparency, Galactic
  extinction, and instrumental configuration. Hence, the background color indicates the fraction of the
  total survey area over which a given limit is reached. White points mark the
  redshift and Ly$\alpha$ luminosities (with error-bars) of the 99
  objects classified as LAEs. The dotted black and white
  lines show the mean and best limits over the whole survey
  respectively. Even below this last limit, the completeness of the
  survey is not zero, explaining why we see 2 points below this curve.}
\label{fig-1}
\end{center}
\end{figure}

The 99 LAEs in the sample span a range in luminosities of ${\rm
log}(L_{Ly\alpha})=42.42-44.03$, and have a median luminosity of 
${\rm log}(\tilde{L}_{Ly\alpha})=43.03$. Figure \ref{fig-1} shows the
survey 5$\sigma$ limiting Ly$\alpha$ luminosity as a function of redshift, together
with the luminosities and redshifts of all LAEs in the sample. The depth
of the observations is variable across the survey area and dependent
on the observing conditions, the airmass at which the observations
were taken, the Galactic dust extinction towards different fields,
and the instrumental configuration.
Colors in Figure \ref{fig-1} correspond to the fraction of the total
surveyed area for which the spectra reaches the corresponding limit in
luminosity. While VIRUS-P has its lower throughput in the blue
end of the wavelength range, the smaller luminosity distance at lower
redshifts compensates for this fact, providing a relatively
flat luminosity limit throughout the entire redshift range. As
mentioned above, detailed
simulations quantifying the completeness and spurious detection ratio
for the whole survey are presented in Paper I. A good
understanding of the completeness of the survey is essential in order
to calculate the Ly$\alpha$ luminosity function. As shown in Paper I,
the completeness at the 5$\sigma$ flux limit shown in Figure
\ref{fig-1} is 33\%, reaching 50\% at 5.5$\sigma$ and 90\% at
7.5$\sigma$.

\begin{figure}[t]
\begin{center}
\plotone{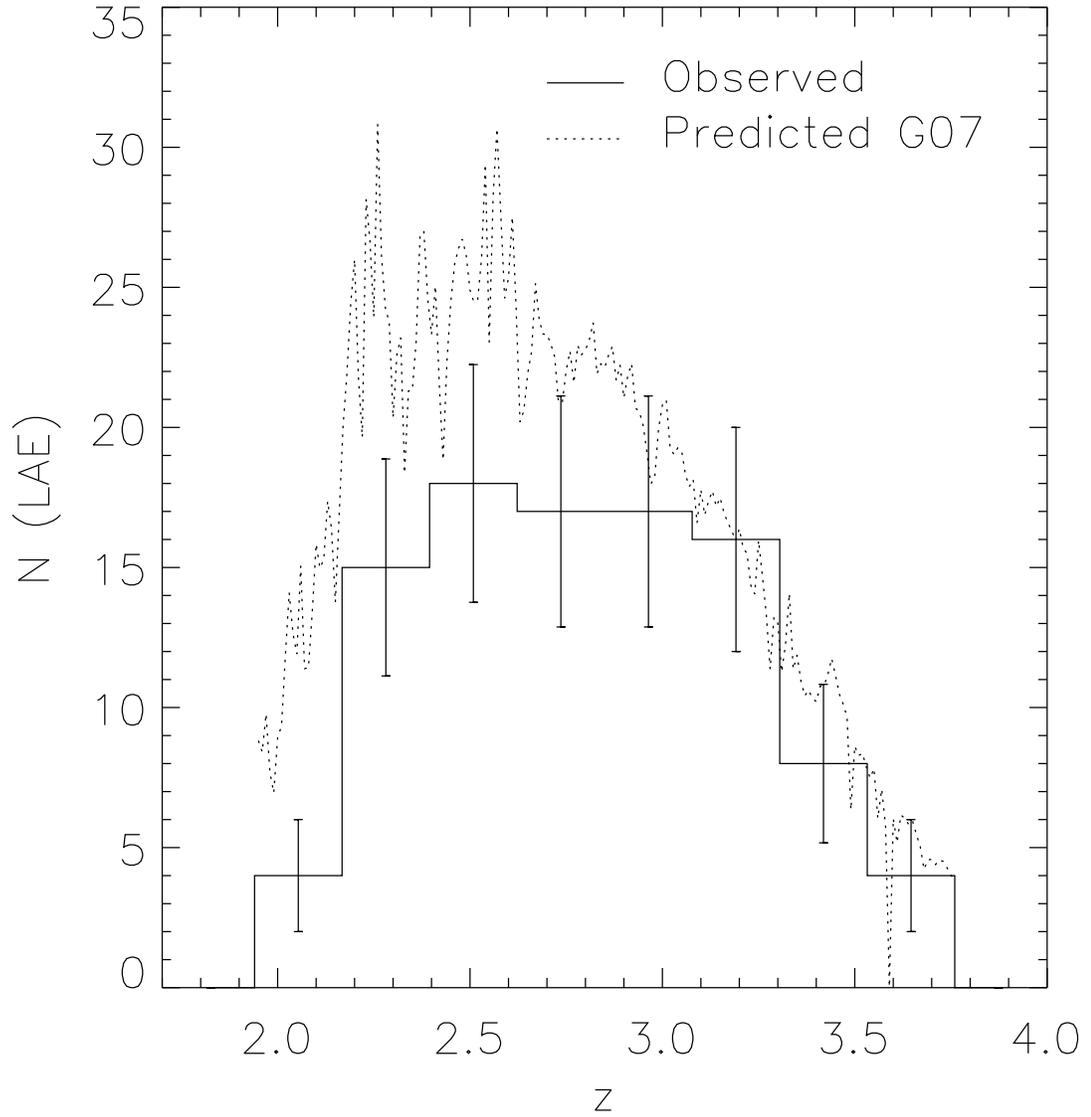}
\caption{Redshift distribution of the 99 LAEs in
  the Pilot Survey (solid histogram). Error-bars represent Poisson
  uncertainties only. Also shown is the incompleteness-corrected predicted
  redshift distribution (dotted line) given by our flux limit and assuming the
  \cite{gronwall07} Ly$\alpha$ luminosity function with no evolution over $2<z<4$.}
\label{fig-2}
\end{center}
\end{figure}

The redshift distribution of LAEs in our sample is shown in Figure
\ref{fig-2} (errorbars show Poisson statistical uncertainties). The
detected galaxies span a range in redshift of
$2.079<z<3.745$, with a median redshift of $\tilde{z}=2.811$, properly sampling
the $1.9<z<3.8$ range over which they could be detected. Figure
\ref{fig-2} also shows the predicted redshift distribution of LAEs in
the Pilot Survey calculated by integrating the \cite{gronwall07}
luminosity function of narrow-band selected LAEs at $z=3.1$ above the Pilot Survey flux
limits shown in Figure \ref{fig-1}, and correcting for the survey
completeness. The agreement is excellent at high redhsift ($z>3$), but we 
observe a drop in the number of LAEs at lower redhsifts from what is predicted by a non-evolving luminosity function. 
Recent narrow-band studies of $z\sim2$ LAEs show
hints for both an increase \citep{guaita10} and
decrease \citep{nilsson09a} of the LAE number density from $z=3$ to
$z=2$. As stated by the authors themselves, neither of these studies probe a
large enough volume to allow for a significant detection of the evolution in the LAE number
density. In our surveyed volume, which is a few times larger than the
volumes surveyed in those studies, we find some evidence for a decrease
in the number density of LAEs from $z\sim4$ to $z\sim2$, although as
discussed in \S6, the statistical uncertainties remain too large to
make a definitive statement. In any case, this type of evolution is
expected if the escape fraction of Ly$\alpha$ photons from galaxies decreases
towards lower redshifts. In \S7 we find evidence that this effect
indeed occurs, which supports the observed drop in the LAE
number density.

\section{The UV slope of Lyman alpha emitters}

The UV continuum slope has been shown to be a powerful tool for
estimating the amount of dust extinction in star-forming galaxies in
the local universe \citep{meurer95, meurer99} as well as at high redshift
\citep{daddi04, bouwens09, reddy10}. Direct observations of the ultra-violet spectral energy
distribution of local star-forming galaxies have demonstrated that in the
1000\AA-3000\AA\ range, they are very well described by a power-law
spectrum of the form $f_{\lambda}\propto \lambda^{\beta}$
\citep{calzetti94}. Differential dust extinction (reddening)
makes the power-law slope correlate well with the
amount of dust extinction in galaxies.

Measuring the spectral slope of the UV SED of LAEs at $1.9<z<3.8$
provides a direct measurement of their dust content, and its
evolution with time. Knowledge of the amount of dust extinction in LAEs
allows us to correct UV measured $SFRs$. An unbiased measurement of the
$SFR$ in these objects is not only important regarding the
star-formation properties of these galaxies, but can also be used,
together with the observed Ly$\alpha$ luminosities, to estimate the
escape fraction of Ly$\alpha$ photons from the ISM of these high
redshift systems, and to study a possible evolution in this quantity.

\subsection{Measurement of the UV continuum Slopes, UV Luminosities,
  and Ly$\alpha$ EWs}

We identify continuum counterparts of spectroscopically detected LAEs
in our sample using publicly available broad-band optical images sampling the
rest-frame UV SED of the objects. Multi-band aperture photometry is
then used to measure their UV continuum slope ($\beta$) and UV
luminosity as described below. 

For the purpose of measuring $\beta$ we use the
B, r$^+$, i$^+$, and z$^+$ images of the COSMOS and GOODS-N fields presented in
\cite{capak04} and \cite{capak07}, the g, r, i, and z images of the
XMM-LSS field from the Canada-France-Hawaii Telescope Legacy Survey
\citep[CFHTLS,][]{mellier08} W1 field, and the g, i, and z MUNICS-Deep images
presented in Paper I. The identification and association with
broad-band counterparts of our emission line detected objects makes
use of a maximum likelihood algorithm which is described in
detail in Paper I. Briefly, our astrometric uncertainty and the
typical surface density of galaxies as a function of continuum brightness is
used to identify the most likely broad-band counterpart for each
LAE. The possibility of the emission-line source having no counterpart
in the broad-band imaging is also considered. This can happen if the source is fainter
than the sensitivity of the images or if the source is spurious. The
no counterpart option is adopted if the probability exceeds that of all other
possible counterparts. Only 9 out of 99 (9\%) objects show no broad-band
counterparts. This number is in good agreement with the 4-10\%
contamination expected from spurious detections in our LAE sample
(Paper I, \S3), although these objects could in principle be real and
have very high $EW$s. In
Paper I we showed that only one of them has significantly high $EW$
given the limits that can be put using the depth of the broad-band
images, while the large majority (8/9) show low signal-to-noise (S/N) detections
($<6.5$) where the false detection ratio is the highest. For
simplicity we omit these ``no counterpart'' sources from our analysis
as we expect the large majority of them to be false detections. We
also reject one other object (HPS-89) from our analysis because its broad-band
counterpart photometry is catastrophically affected by a bright neighbor.

Fluxes are measured in optimal $1.4\times$FWHM
diameter color apertures, and scaled to total fluxes for each object
using the ratio between V-band (g-band for MUNICS) fluxes measured in
the color aperture and aperture-corrected fluxes measured in a
SExtractor \citep{bertin96} defined Kron aperture \citep{gawiser06a,blanc08}. Any
contribution from the measured Ly-alpha line to the broad-band fluxes is
removed. While we do take into account IGM absorption when fitting for the UV slope, we
decide to omit the U-band in the fits because in our redshift
range the band includes the Lyman 912\AA\ break. Since IGM absorption is
expected to be stochastic, an average line-of-sight correction
might not apply to single objects. This leaves us with a B-band
through z-band SED for each object.

\begin{figure}[t]
\begin{center}
\plotone{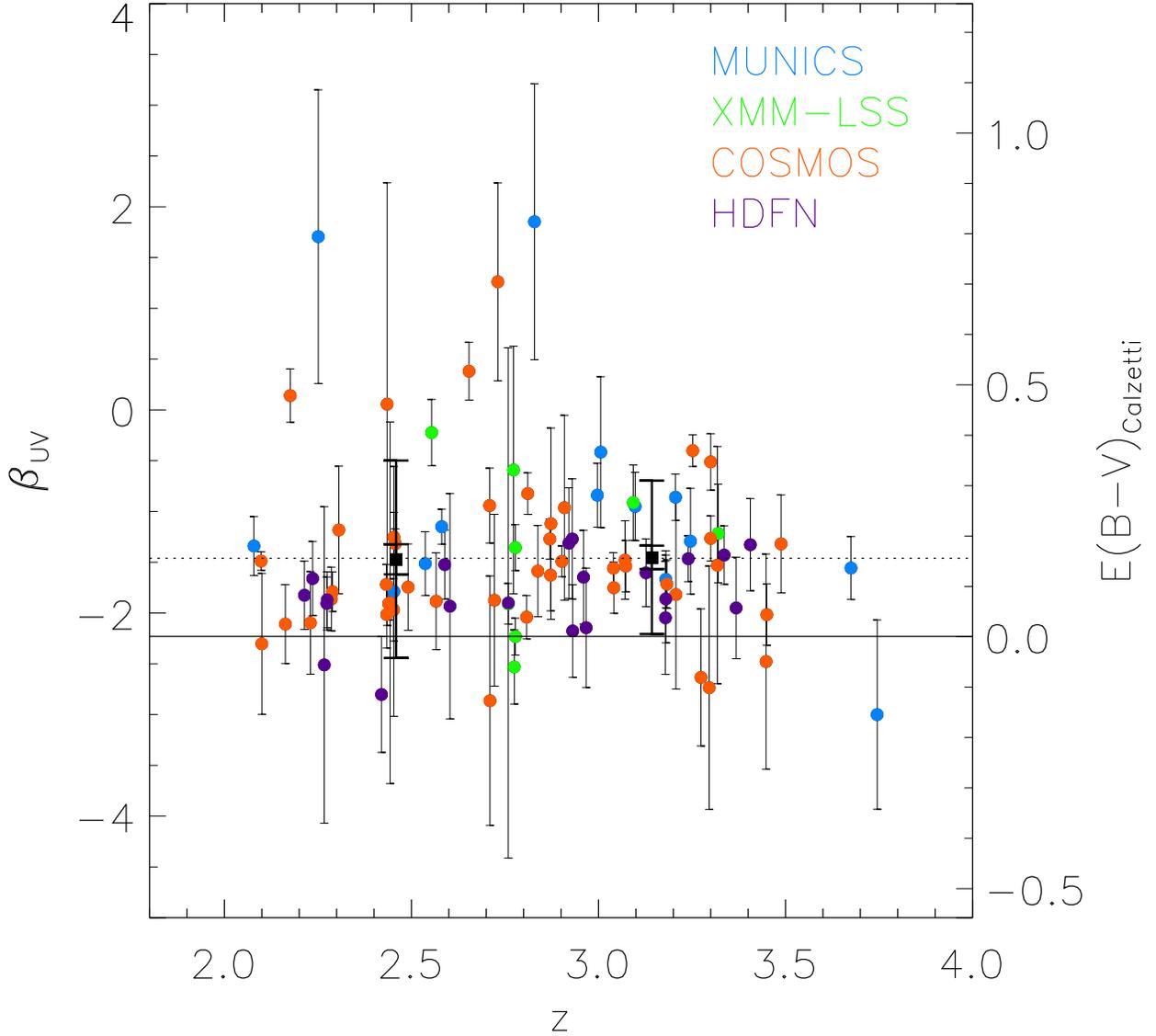}
\caption{UV continuum slope as a function of redshift for the 89 LAEs
  with broad-band optical counterparts. Objects are color coded by
  field. The right axis shows the equivalent E(B-V) assuming a
  \cite{calzetti00} attenuation law. The horizontal lines mark the
  assumed intrinsic UV slope corresponding to a dust-free stellar
  population ($\beta_0=-2.23$, solid line), and the mean for the whole sample (dotted
  line). Also shown are the
  mean UV slopes for two redshift bins at $1.9<z<2.8$ and
  $2.8<z<3.8$ (black squares), with two sets of error-bars corresponding to the
  standard deviation in $\beta$ within each bin (large error-bars) and
  the formal error in the mean (small error-bars).}
\label{fig-3}
\end{center}
\end{figure}

The approximate rest-frame wavelength range sampled by the above bands
shifts from 1500\AA-3000\AA\ at $z=1.9$ to 900\AA-1900\AA\ at
$z=3.8$, so only the B-band is affected by Lyman forest absorption at
the higher redshift end of our range. Following a similar
methodology as that described in \cite{meurer99} and
\cite{reddy10}, we compute the UV continuum slope for each object by fitting the
rest-frame UV SED with a
power-law spectrum of the form $f_{\lambda}\propto \lambda^{\beta}$
corrected for IGM absorption at the corresponding redshift of each
object using a \cite{madau95} prescription. All available
bands redwards the Lyman break are used to perform the fit. When an object is not detected in a
particular band, we properly include the upper limit in flux given by
the photometric uncertainty in the $\chi^2$ minimization in order to
not censor our data. The error in $\beta$ is estimated from Monte Carlo
simulations of 100 realizations of the UV SED, where the fluxes in each band are varied
within their photometric errors. This fitting also provides the
UV luminosities at 1216\AA\ and 1500\AA\ which are used to
estimate the Ly$\alpha$ $EW$ and the $SFR$, respectively. All these
quantities are reported in Table \ref{tbl-1}. Figure
\ref{fig-3} shows the UV continuum slope $\beta$ as a function
of redshift for the 89 objects having continuum counterparts.

In principle, the UV slope can depend not only on the amount of dust extinction, but also on the
age, metallicity, and initial mass function (IMF) giving rise to the stellar
population. Extensive work can be found in the literature
regarding these effects on the observed UV slope of star-forming
galaxies. \cite{leitherer95} showed that for both
instantaneous bursts and constant
star formation synthetic stellar populations, changes of the
order of $\Delta \beta=\pm0.2$ around a typical value of
$\beta \sim-2.3$ are introduced by variations in age
(1 Myr to 1 Gyr) and metallicity ($0.1\;Z_{\odot}$ to $2\;Z_{\odot}$). They
also find the UV slope to be largely insensitive to the assumed
IMF. This result is in good agreement with the work of \cite{bouwens09}, 
who demonstrate that the UV slope dependence on
dust is dominant over that on age, metallicty and IMF. They use
\cite{bruzual03} models to show that changes by a factor of two in age
and metallicity introduce changes of $\Delta\beta\lesssim0.1$. \cite{schaerer05}
also present a similar result. In Figure 1 of their paper it can be
seen that for a range in ages of 1 Myr to 1 Gyr (encompassing the
expected age range for LAEs), and metallicities between 1/50
Z$_{\odot}$ and solar, both constant SFR population synthesis models,
and single bursts younger than 10 Myr (time over which they can
produce significant Ly$\alpha$ emission) show variations in their UV
slopes of $\Delta \beta=\pm0.2$  ($\Delta E(B-V)=0.04$). These
systematics are smaller than the typical uncertainty in the
measurement of $\beta$ for LAEs in our sample. Therefore, by assuming
a constant value for the intrinsic (dust-free) UV slope across our
LAE sample, we can robustly estimate the amount of dust reddening
directly from the observed values of $\beta$ given an attenuation
law. 

The right axis of
Figure \ref{fig-3} shows the corresponding value of the reddening
$E(B-V)$, calculated assuming an 
intrinsic UV slope $\beta_0=-2.23$ for a dust-free stellar population
\citep{meurer99}, and a \cite{calzetti00} extinction law. The value
of $\beta_0$ is derived from a fit to the relation between the IR to
UV ratio and $\beta$ in a sample of local starburst galaxies
\citep{meurer99}, and reproduces the observed
relation at $z\sim2$ \citep{reddy10}. Although \cite{reddy10} found
young ($<100$ Myr) $z\sim2$ galaxies to lie slightly below the \cite{meurer99}
relation, and closer to that of \cite{pettini98}, these two
relations converge at low extinction and imply basically
indistinguishable values for $\beta_0$. In order to take into account
age and metallicity induced uncertainties in our error budget for the
dust reddening, we sum in quadrature
a systematic error of $\Delta\beta=\pm0.2$ ($\Delta E(B-V)=0.04$) to
the uncertainty in $\beta$, and propagate it into the error in
$E(B-V)$. Measured values for the dust reddening and it associated
uncertainty are reported in Table \ref{tbl-1}.

\subsection{Dust Properties of LAEs and comparison to Previous Measurements}

Our LAEs show a mean UV continuum slope $\langle \beta
\rangle=-1.5\pm0.1$ (formal error on the mean) corresponding to a mean $\langle E(B-V)
\rangle=0.16\pm0.02$ (median $\tilde{E}(B-V)=0.13$). The measured slopes span a relatively broad range of
$-3<\beta <+2$, with the large majority (83/89, 93\%) of the objects having
$\beta<0$ ($E(B-V)<0.45$). All objects with $\beta <
\beta_0$ (i.e. bluer than the assumed intrinsic dust-free slope) are
consistent with $\beta=\beta_0$ (i.e. $E(B-V)=0$) within 1$\sigma$.

These slopes and reddenings are in rough agreement with previous
measurements of narrow-band detected LAE broad-band
colors. For $z=2.1$ LAEs, \cite{guaita10} find a typical
$(B-R)\simeq0.2$ ($\beta=-1.5$ 
using equation 3 in \cite{nilsson09a}) and a relatively uniform
distribution in the $-0.5<(B-R)<1$ ($-3.3<\beta<0.7$)
range. Similarly, at $z=2.3$, \cite{nilsson09a} find a median
$(B-V)=0.14$ corresponding to $\beta=-1.4$, with the bulk of their LAEs
having $-3.0 < \beta < 2.0$. At $z=2.2$ \cite{hayes10a} used SED fitting
to find that LAEs in their sample have a range in $E(B-V)=0-0.4$. 
At higher redshifts, usually lower levels of extinction are
measured. At $z=3.1$ \cite{nilsson07} find $A_{V}=0.26^{+0.11}_{-0.17}$ from
fitting the stacked SED of 23 LAEs in the GOODS-S field, corresponding to
$E(B-V)=0.06^{+0.03}_{-0.04}$ (assuming a Calzetti attenuation
law). \cite{verhamme08}, using Monte Carlo
Ly$\alpha$ radiative transfer fitting of the line profiles of 11
$z\sim 3-5$ LBGs (8 of them also LAEs) from \cite{tapken07}, find that
the color excess spans a range of $E(B-V)=0.05-0.2$. \cite{gawiser06a}
report that the best-fit SED to the
stacked optical photometry of $z=3.1$ LAEs in their sample has
$A_V=0^{+0.1}_{-0.0}$, corresponding to $E(B-V)<0.03$.

\begin{figure}[t]
\begin{center}
\plotone{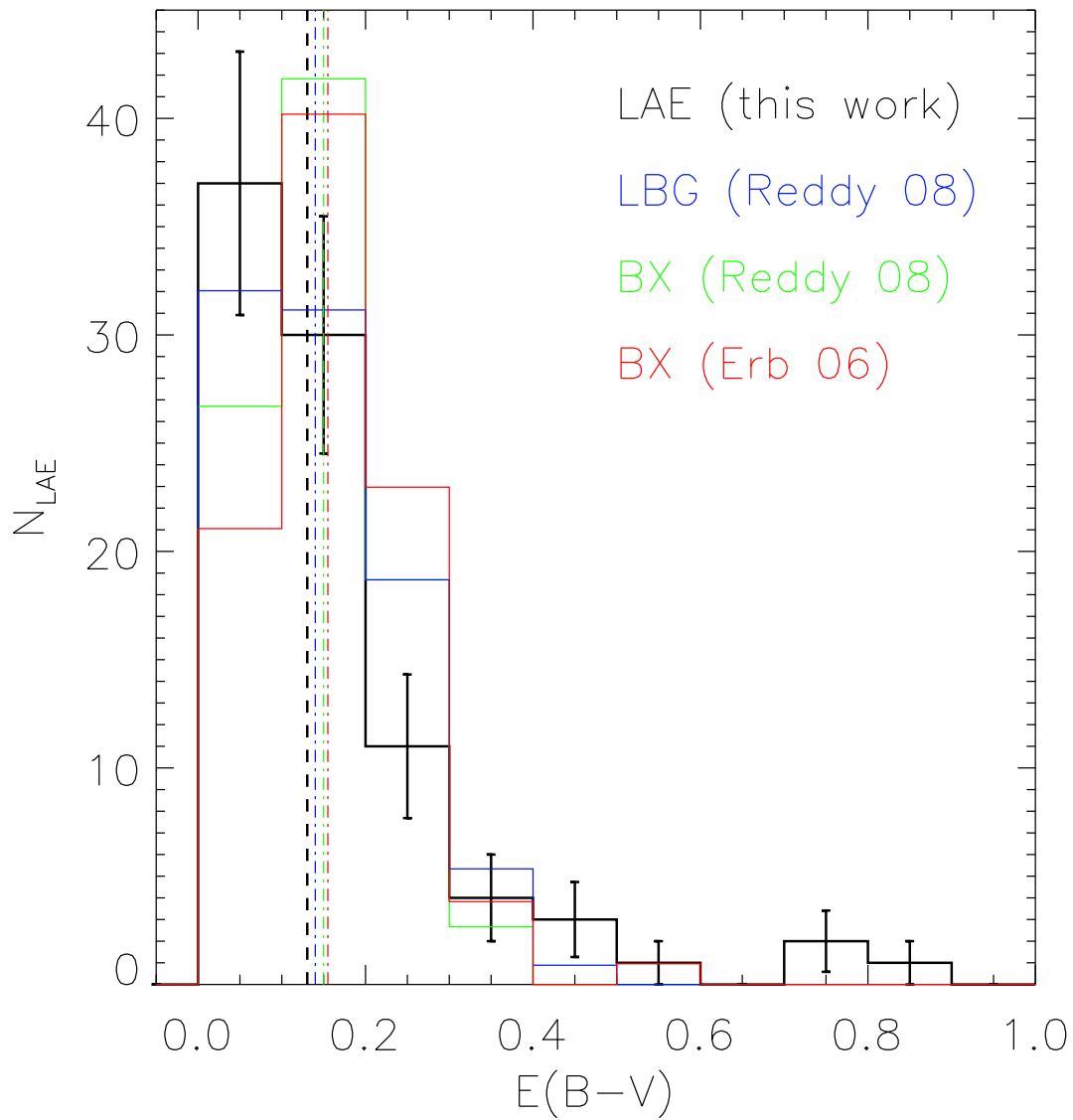}
\caption{$E(B-V)$ distribution of LAEs in our sample (Poisson error-bars), together with that
of BX/LBGs taken from \cite{erb06} and \cite{reddy08} (solid
histograms). The median of each distribution is marked by the vertical
dashed lines.}
\label{fig-4}
\end{center}
\end{figure}

For comparison, similar ranges in $\beta$ and $E(B-V)$ as those
seen here have been measured for
LBGs \citep[e.g.][]{shapley01, erb06, reddy08}. Figure \ref{fig-4}
shows the $E(B-V)$ distribution of LAEs in our sample, compared with
that of UV continuum selected galaxies at $1.9<z<2.7$ (BX galaxies)
and at $2.7<z<3.4$ (LBGs) from
\cite{erb06} and \cite{reddy08}. The $E(B-V)$ distributions for the
continuum-selected galaxies
are different from those presented in the original papers in that we have
set all their $E(B-V)<0$ values to zero for proper comparison with our
sample. It can be seen that both the shape and median value of
the $E(B-V)$ distribution of LAEs and BX/LBGs are relatively similar (medians
are indicated by the dashed lines in Figure \ref{fig-4}). This result, together
with the fact that LAEs and BX/LBGs
seem to overlap in the two-color BX/LBG selection diagram
\citep{guaita10, gawiser06a}, implies that both populations have
relatively similar spectral continuum properties in the
UV. Nevertheless, Figure \ref{fig-4} shows an LAE distribution that is
peaked at lower $E(B-V)$ than the BX/LBG distributions, and also that
Ly$\alpha$ selection might allow for the inclusion of some highly reddened
objects, although the reality of these red LAEs will be questioned in
the next section. These galaxies, if real, are excluded of UV-selected samples by
construction, since the color cuts in those selections reject object
with $E(B-V)\gtrsim 0.5$ \citep{daddi04, blanc08}.

The observed UV slopes imply that LAEs present low levels of dust
extinction. One third (30/89) of the LAEs in our sample are
consistent with being dust-free ($E(B-V)=0$) to 1$\sigma$, with the
fraction going up to 60\% within the 2$\sigma$ uncertainty. Still, a
significant fraction of LAEs show non negligible amounts of dust. As
will be shown in \S5, dust in LAEs should not be neglected;
doing so would strongly underestimate the $SFR$ in these
objects. Dust also plays a dominant role in setting the escape
fraction of Ly$\alpha$ photons, as we will discuss in \S5.4.

\subsection{Evolution of the Dust Properties of LAEs}

At first sight, Figure \ref{fig-3} shows different behaviors in the
dust-content distribution of LAEs at the high and low redshift ends of
our sample. At $z<3$ we see the emergence of a small population of
LAEs (6/89) with red UV slopes ($\beta>0$). This objects, if real, could
represent an interesting population of dusty star forming galaxies in
which some physical mechanisms allows for the escape of Ly$\alpha$
photons. We have reasons to question the reality of these objects (see
below). Furthermore, in this section we show that their presence
does not affect the average properties of the LAE population which is
dominated by UV-blue LAEs.

To test for any evolution on the dust-content of LAEs with redshift we
divided the sample in two redshift bins: low-$z$ ($1.9<z\le2.8$) and high-$z$
($2.8<z<3.8$). The division corresponds to the median redshift of the
whole sample, and divides the survey volume in two roughly
equal sub-volumes. The corresponding age of the universe at $z=3.8$, 2.8,
and 1.9 is $\sim$1.6, 2.3, and 3.4 Gyr. The median Ly$\alpha$
luminosity of the two sub-samples equals that of the whole sample
(log$L_{Ly\alpha}=$43.0). 

The mean UV slopes of the high and low redshift
samples are shown in Figure \ref{fig-3}.
Error-bars show the formal error on the mean and the standard
deviation for each sample including
$\sim$68\% of the objects. The mean value of $E(B-V)$
stays constant with values $0.16\pm0.03$ and
$0.17\pm0.02$ for the low and high redshift bins respectively. The
scatter around these
values is large and the means are statistically consistent with each
other, and with the mean of the full sample. Therefore, we do not
detect any significant evolution in the average UV slope and dust
reddening of LAEs from $z\sim4$ to $z\sim2$.

The lack of evolution in the mean dust-content of LAEs implies that
this rare population of very high $E(B-V)$ objects emerging at
$z<3$, if real, does not affect the average
properties of the overall population due to their reduced number. The
dust content of the bulk of the LAE population
remains relatively constant across the $2<z<4$ range. 

\begin{figure}[t]
\begin{center}
\plotone{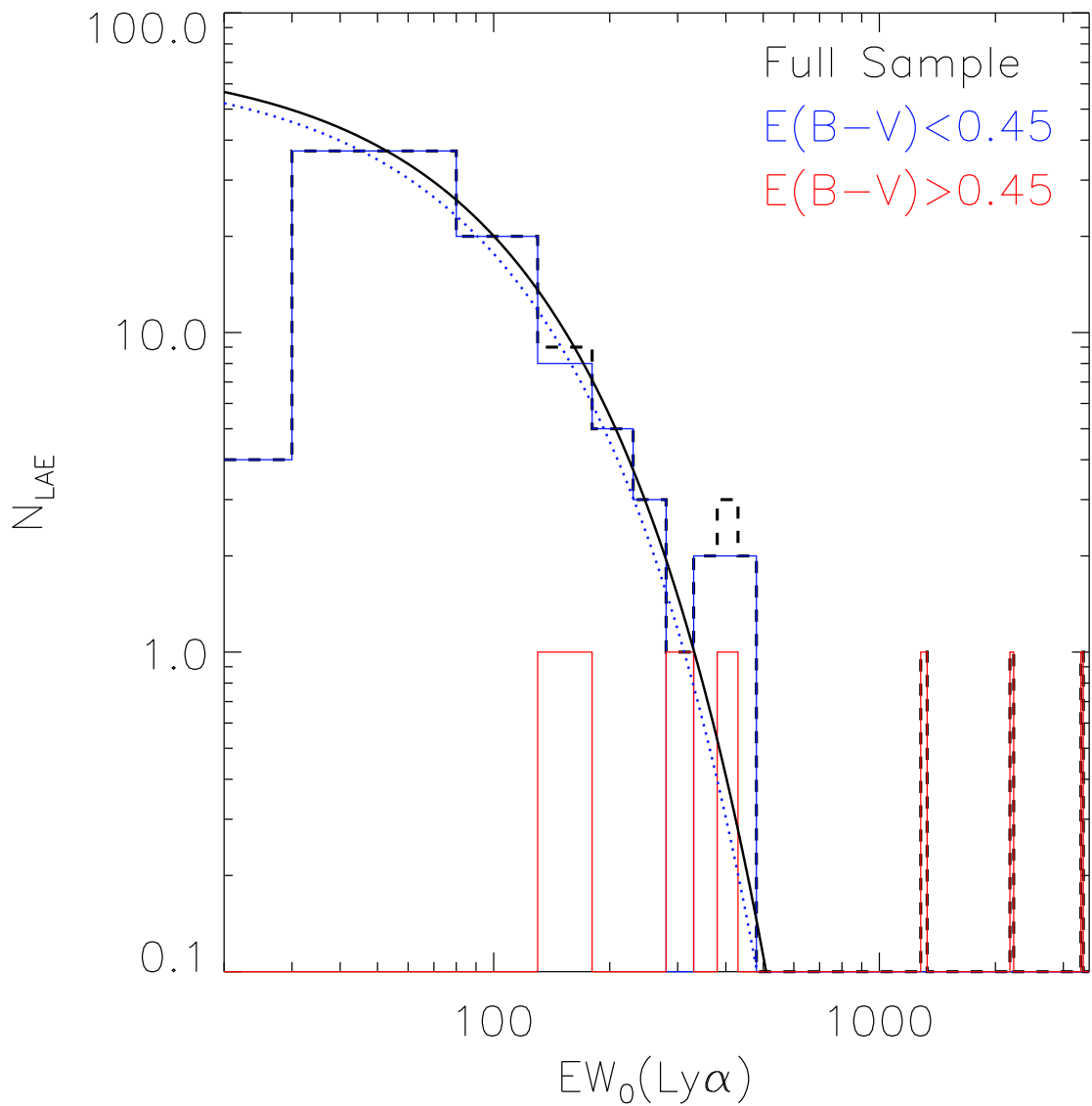}
\caption{Rest-frame Ly$\alpha$ $EW$ distribution of LAEs in our sample
  (dashed black histogram). The distributions for low
  ($E(B-V)<0.45$) and high ($E(B-V)>0.45$) reddening objects are shown
(blue and red histograms respectively). Also shown are the
  best-fit exponential distribution ($N\propto \exp{[-EW/w_{0}]}$) to
  the whole sample ($w_0=77\pm7$\AA, solid black line) and the low
  reddening sample ($w_0=74\pm7$\AA, dotted blue line).}
\label{fig-5}
\end{center}
\end{figure}

Doubt regarding the validity of the UV slope measurements for these
objects, and their classification as LAEs arises from looking at the
distribution of rest-frame Ly$\alpha$ $EWs$ for our sample. Figure \ref{fig-5}
shows the $EW$ distribution for both UV-blue
($\beta<0$) and UV-red ($\beta\geq 0$) LAEs together with
the one for the whole sample. The Ly$\alpha$ $EW$ is measured as
described in \S4.1, and hence can differ from the values presented in
Paper I. It is evident from Figure \ref{fig-5} that UV-blue LAEs
dominate the overall population since they present a practically
indistinguishable $EW$ distribution (well fitted by an exponential
with an e-folding parameter $w_{0}=74\pm 7$) from that of the full
sample ($w_{0}=77\pm 7$). UV-red LAEs on the other hand, in addition to
being rare in numbers, present a very different distribution in
rest-frame $EW$, characterized by the presence of many extremely high $EW$
($>500$\AA) objects. Two of these UV-red objects are in the MUNICS
field where we lack deep X-ray data to reject AGNs from our
sample (one of these sources shows significantly extended Ly$\alpha$ emission
and is a good candidate for an extended Ly$\alpha$ nebula, or
Ly$\alpha$ Blob as discussed in Paper I). The
remaining four objects have low association probabilities ($\leq0.6$)
with their broad-band counterparts, casting doubt on the validity of
our UV slope and $EW$ measurements for these objects. Further
follow-up observations are necessary to confirm the nature of these detections.

If real, these UV-red LAEs would have an extreme nature, being very
dusty and highly star-forming. We remove these six objects from all the
subsequent analysis, and for the rest of the paper we focus only on the
results regarding the dominant UV-blue LAE population. It must be kept
in mind that if these objects happen to be real LAEs, no strong evidence
for a bimodality in the dust content or $SFR$ of LAEs is found in our
data. The $\beta>0$ cut used to separate this population of objects is
solely based on the fact that $\beta>0$ objects are absent at $z>3$ in
our sample. After removing these objects from our sample, we find a
mean dust-reddening for LAEs of $\langle E(B-V) \rangle=0.13\pm0.01$,
corresponding to an average dust attenuation of $\sim70$\% at 1500\AA.

\section{UV versus Ly$\alpha$ $SFR$\lowercase{s} and the Escape Fraction of Ly$\alpha$ Photons}

In this section we use the dust extinction values derived from the
UV continuum slope in the previous section to estimate the
dust-corrected $SFR$ of LAEs in our sample. A comparison between the
observed Ly$\alpha$ luminosity and the intrinsic Ly$\alpha$ luminosity
implied by the dust-corrected $SFR$ allows estimation of the escape
fraction of Ly$\alpha$
photons from these galaxies. Throughout this analysis we have decided to
neglect the effects of the IGM. As stated in \S1, at these redshifts we
expect attenuations for Ly$\alpha$ of no more than
5-25\%, which is within our typical uncertainty for the Ly$\alpha$
luminosity. Furthermore, if outflows are common in LAEs, as many lines
of evidence suggest, then IGM scattering at these redshifts may become
even less important as most Ly$\alpha$ photons leave galaxies red-shifted from the
resonance wavelength (see discussion and references in \S1). We start
by comparing the observed (not
corrected for dust) $SFRs$ derived from the UV and Ly$\alpha$
luminosities, then introduce the dust corrections, and finally estimate the
escape fraction of Ly$\alpha$ photons and study how it relates to the amount
of dust reddening.

\subsection{Estimation of the Star Formation Rate and the observed
  $SFR(Ly\alpha)$ to $SFR(UV)$ ratio.}

The UV monochromatic luminosity at 1500\AA\ ($L_{\nu,1500}$) for each object is taken
from the fits described in \S4. In order to calculate the $SFR$ we use a
standard \cite{kennicutt98} conversion

\begin{equation}
SFR(UV)\;[{\rm M_{\odot} yr^{-1}}] =1.4\times 10^{-28} L_{\nu,1500} [\rm{erg\;s^{-1} Hz^{-1}}]
\end{equation}

\noindent
which assumes a Salpeter IMF with mass limits 0.1 to 100
M$_{\odot}$. The Ly$\alpha$ derived $SFRs$ were calculated using the
standard \cite{kennicutt98} conversion factor for H$\alpha$ and assuming the
intrinsic Ly$\alpha$ to H$\alpha$ ratio of 8.7 from Case B recombination
theory \citep{brocklehurst71, osterbrock06}, so

\begin{equation}
SFR(Ly\alpha)\;[{\rm M_{\odot} yr^{-1}}] =7.9\times 10^{-42} \frac{L_{Ly\alpha}}{8.7} [\rm{erg\;s^{-1}}]
\end{equation}

\begin{figure}[t]
\begin{center}
\plotone{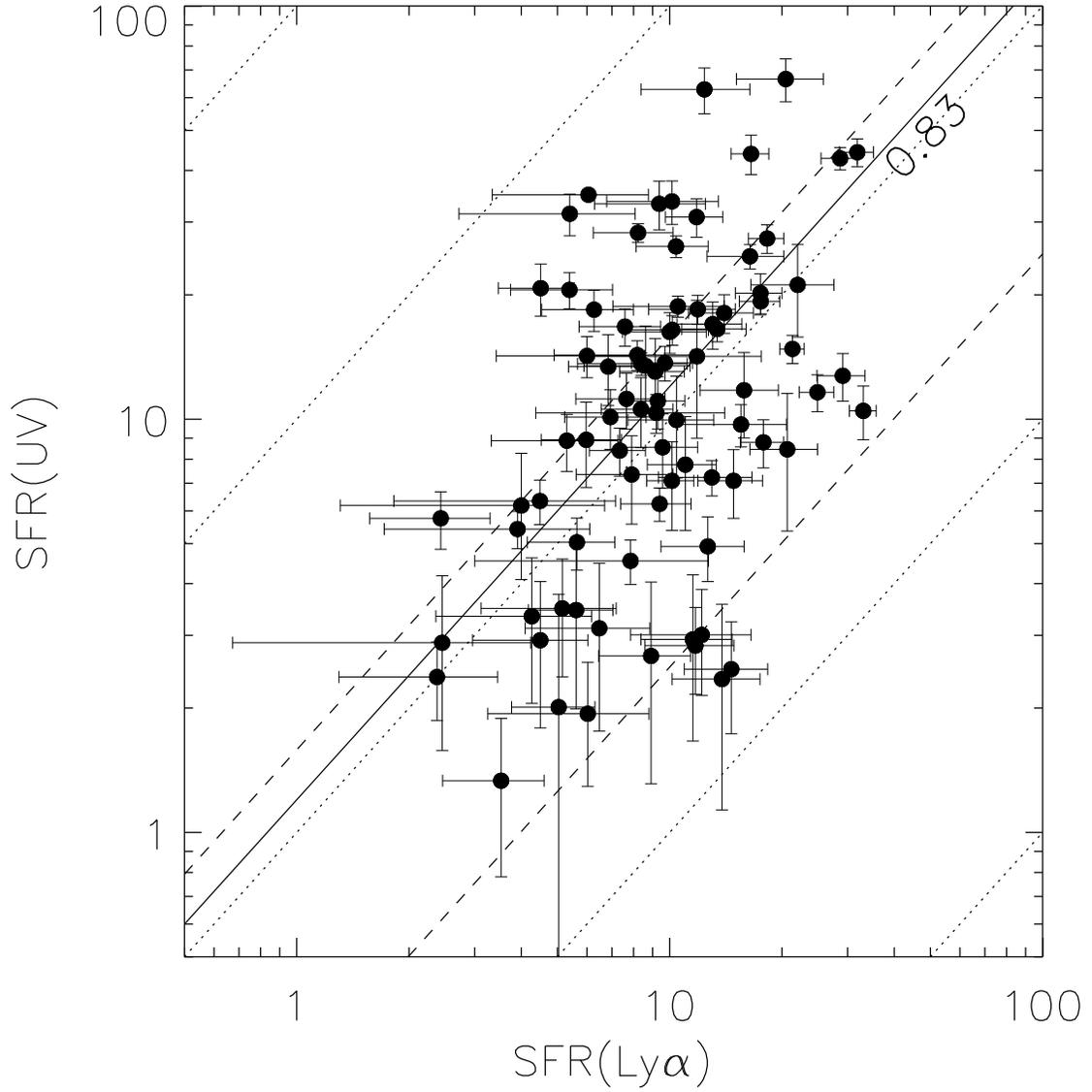}
\caption{UV versus Ly$\alpha$ derived $SFRs$ for the 83 LAEs in the
  final sample. Values
  are not corrected  for dust extinction. The solid line shows the
  median $SFR$(Ly$\alpha$) to $SFR(UV)$ ratio of 0.83. The expected range
  for dust-free normal stellar populations is marked by the dashed
  lines. Dotted lines mark ratios of 0.01, 0.1, 1, 10, and 100.}
\label{fig-6}
\end{center}
\end{figure}

Figure \ref{fig-6} shows $SFR(Ly\alpha)$ versus $SFR(UV)$ for our 83 objects. 
Without accounting for dust we measure median $SFRs$ of 11
M$_{\odot}$yr$^{-1}$ and 10 M$_{\odot}$yr$^{-1}$ from UV continuum
and Ly$\alpha$ respectively. Although these agree with what is typically quoted for
LAEs in the literature, we consider them to be underestimated by
roughly a factor of $\sim3-4$ because of the lack of a dust extinction
correction.

We observe a median ratio $SFR(Ly\alpha)/SFR(UV)=0.83$. Single objects
present a large scatter around the median, with values ranging from 0.2
to 5.9. Since the UV $SFR$ conversion factor is valid
for galaxies with constant star formation over $100$ Myr or more, while
the one for Ly$\alpha$ is valid at much younger ages of $\sim10$ Myr
\citep{kennicutt98}, young galaxies can have intrinsic $SFR(Ly\alpha)$ to
$SFR(UV)$ ratios higher than unity. The dashed lines in Figure
\ref{fig-6} show the allowed range for dust-free constant
star-formation stellar populations with metallicities from 1/50
Z$_{\odot}$ to solar and ages from 1 Myr to 1 Gyr from
\cite{schaerer03}. A Ly$\alpha$ escape fraction
of less than unity can push objects above this range. All the objects
in our sample show $SFR(Ly\alpha)$ to $SFR(UV)$ ratios (or roughly
equivalently Ly$\alpha$ $EW$s), which are consistent within 1$\sigma$ with those of
normal stellar populations (i.e lower than $\sim4$). 

The observed median ratio between these two
quantities is in rough agreement with previous measurements
found in the literature. At $z\sim2.1$, \cite{guaita10} measures a mean
$SFR(Ly\alpha)$ to $SFR(UV)$ ratio of 0.66 for narrow-band selected
LAEs, consistent with the 0.56 value measured by \cite{nilsson09a} at
$z\sim 2.3$. \cite{gronwall07} find LAEs at $z\sim 3.1$ to span a
similar range in the
$SFR(Ly\alpha)$-$SFR(UV)$ plane as the one observed here, and while
they quote a mean ratio of
0.33, a revised value of $\sim 1$ is actually a better estimate for
their sample \footnote{Caryl Gronwall, private communication}. \cite{ouchi08} measures
a ratio of 1.2 in their $z\simeq 3.1$ sample of LAEs. Recently
\cite{dijkstra10} conducted a statistical study of the relation
between these two quantities. Compiling a number of LAE samples
at $3.0\lesssim z \lesssim 6.5$, they find 68\% of LAEs to show
$SFR(Ly\alpha)/SFR(UV)=0.9^{+1.6}_{-0.5}$, in agreement with
our observations.

\begin{figure}[t]
\begin{center}
\epsscale{0.6}
\plotone{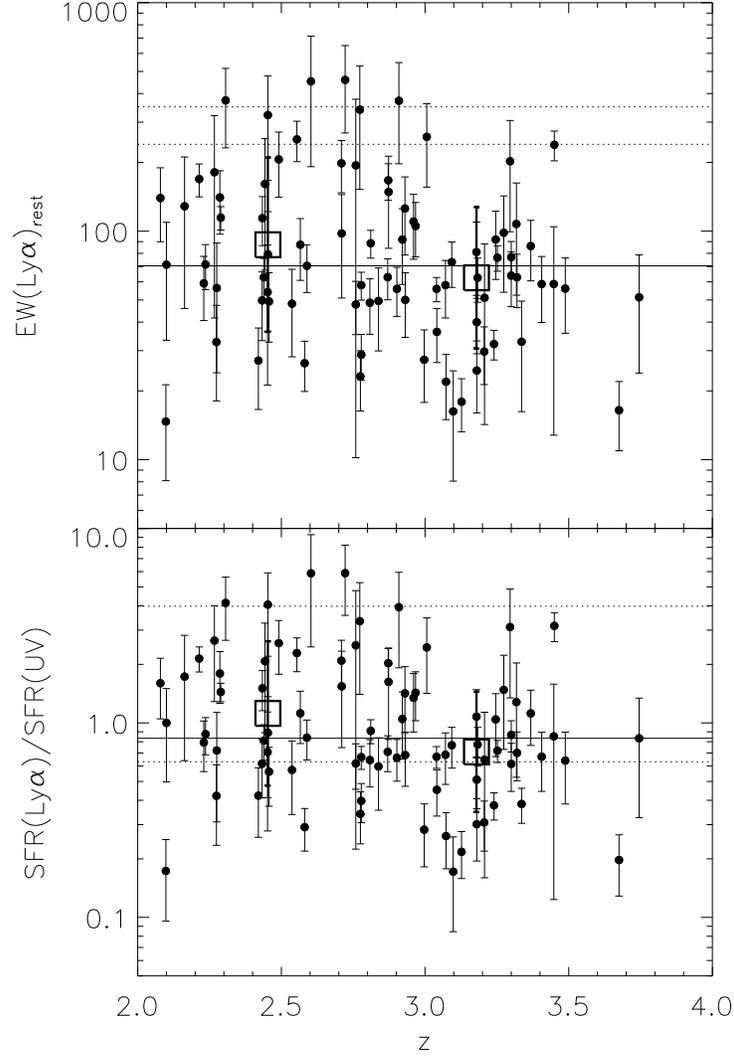}
\caption{Rest-frame Ly$\alpha$ $EW$, and $SFR(Ly\alpha)$ to $SFR(UV)$
  ratio (not corrected for dust) as a
 function of redshift. The median $EW$ of 71\AA\, and ratio of 0.83
 are marked by solid horizontal lines. The dotted lines on the top
 panel indicate the maximum $EW$ range for young normal stellar
 populations with  metallicities between solar and one 1/50 solar from
 \cite{schaerer03}. Dotted lines in the bottom panel display the allowed
 range in the $SFR(Ly\alpha)$ to $SFR(UV)$
  ratio for dust-free normal stellar populations. The open boxes show the
  median $EW$ and ratio for
 the two redshift bins at $1.9<z<2.8$ and $2.8<z<3.8$.}
\label{fig-7}
\end{center}
\end{figure}

\begin{figure}[t]
\begin{center}
\epsscale{0.6}
\plotone{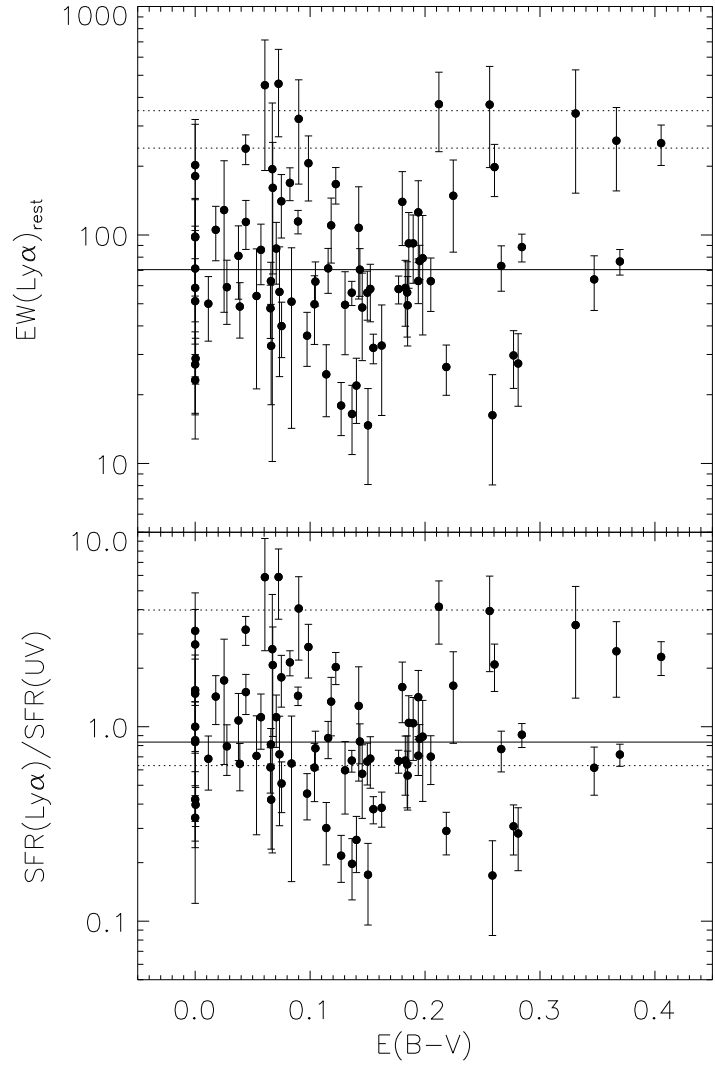}
\caption{Rest-frame Ly$\alpha$ $EW$ and $SFR(Ly\alpha)$ to $SFR(UV)$
  ratio (not corrected for dust) as a function of E(B-V). Symbols are
  the same as in Figure \ref{fig-7}.}
\label{fig-8}
\end{center}
\end{figure}

There is reason to expect evolution in $SFR(Ly\alpha)/SFR(UV)$
with redshift. First, if the dust content of galaxies changes
with redshift, and Ly$\alpha$ and UV photons suffer different amounts
of extinction, we should see a redshift dependence in the ratio. Also,
if the Ly$\alpha$ line suffer from significant IGM absorption, the
dependence of the IGM opacity with redshift should
affect the $SFR(Ly\alpha)$ to $SFR(UV)$ ratio. In Figure \ref{fig-7} we
present the $SFR(Ly\alpha)$ to $SFR(UV)$ ratio, as well as the rest-frame
Ly$\alpha$ $EW$ as a function of redshift. While these two quantities
are roughly equivalent, $SFR(UV)$ is calculated from the UV
monochromatic luminosity
at 1500\AA, while the $EW$ uses the monochromatic luminosity at 1216\AA, therefore
the ratio between them has a mild dependence on the UV
slope. Because of this dependence, we chose to present both quantities in Figures
\ref{fig-7} and \ref{fig-8}.

Over the $2<z<4$ range, we do not observe evolution at a significant
level in the Ly$\alpha$ $EW$ or the ratio between the
Ly$\alpha$ and UV $SFR$s. For our low and high redshift bins we measure
median $EW$s of $87\pm63$\AA\ and $53\pm26$\AA\ respectively (median
absolute deviation errors). A Kolgomorov Smirnof (KS) test to the cumulative $EW$
distributions for the low and high redshift sub-samples 
allows the hypothesis of them being drawn from the same parent
distribution to 2$\sigma$. In terms of $SFR(Ly\alpha)/SFR(UV)$, the
measured median ratios are 1.1 and 0.7 for the low and high redshift
sub-samples. The fact that we do not observe a significant decrease in the
typical $EW$ of LAEs supports our assumption of neglecting IGM
absorption in our analysis.

We also analyze the relation between $SFR(Ly\alpha)/SFR(UV)$ and the
dust reddening $E(B-V)$ derived from the UV slope. If UV and
Ly$\alpha$ photons suffer from similar amounts of extinction, the
above ratio should be independent of the amount of dust present in the
galaxy. This is indeed the case for our LAEs, as can be seen
in Figure \ref{fig-8}, where the relation for the two quantities (as
well as that between $EW$ and $E(B-V)$) is shown. Throughout the entire
range $0<E(B-V)<0.45$ the
ratio between Ly$\alpha$ and UV derived $SFRs$ stays flat with objects
scattered around the median value. A similar behavior is seen for the
$EW$.

\subsection{Dust Corrected $SFRs$ and Estimation of the Ly$\alpha$
  Escape Fraction}

\begin{figure}[t]
\begin{center}
\plotone{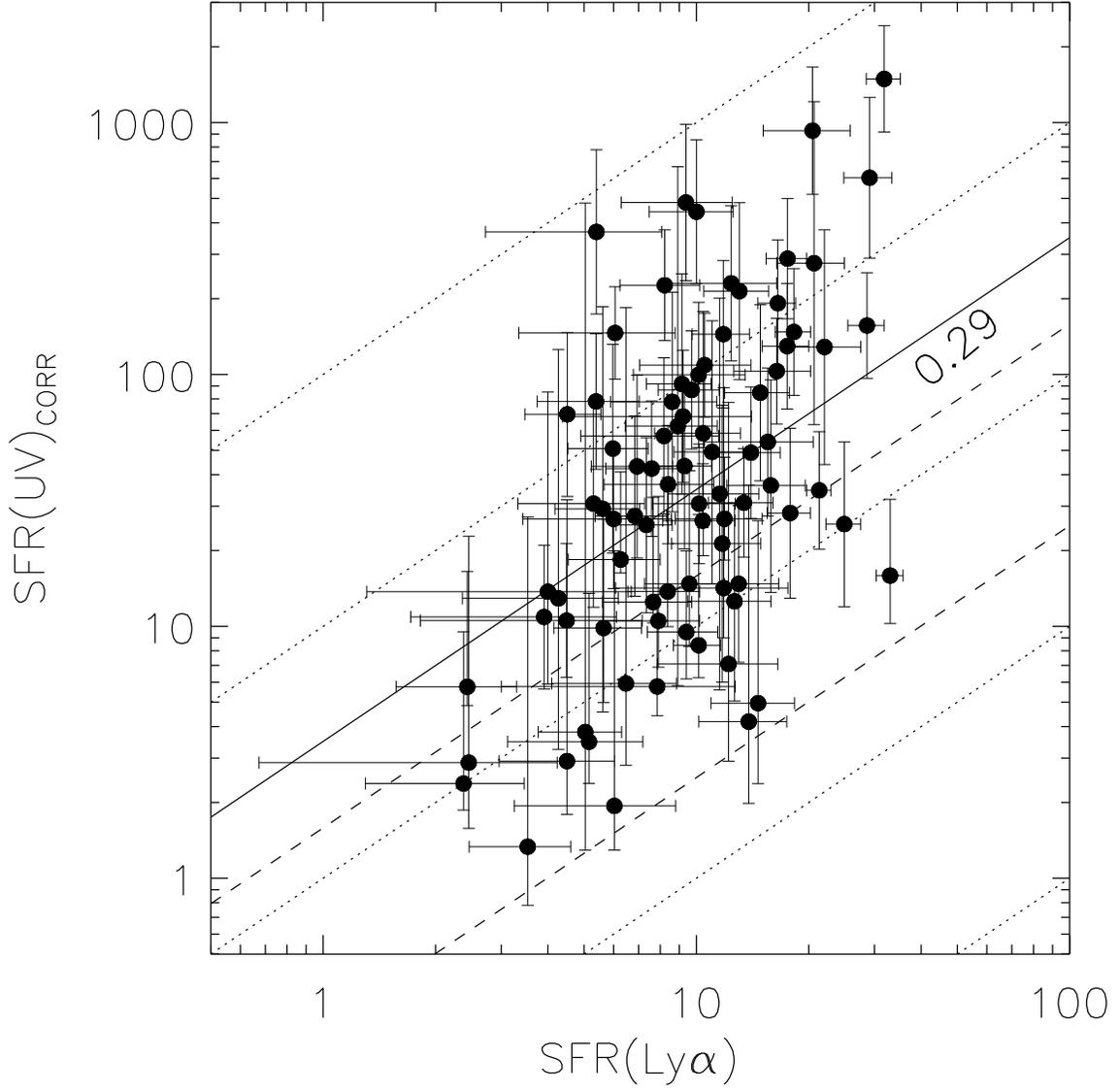}
\caption{Same as Figure \ref{fig-6}, but with $SFR(UV)$ corrected for
  dust. Error-bars include the uncertainty in the correction. The
  solid line marks the median escape fraction of 29\%.}
\label{fig-9}
\end{center}
\end{figure}

We now correct the UV luminosity of our objects using the values of
$E(B-V)$ estimated in \S4 and a \cite{calzetti00} attenuation
law. This approach provides a better estimate of the true $SFR$ in the
galaxies. Figure \ref{fig-9} shows a comparison between the
dust-corrected $SFR(UV)_{corr}=SFR(UV)\times10^{(0.4k_{1500}E(B-V))}$,
and the uncorrected $SFR(Ly\alpha)$. Error-bars include the
uncertainty in the dust correction which has been propagated from the
uncertainty in the measurement of the UV continuum slope $\beta$. Note
that the axes in Figure \ref{fig-9} are different from those in Figure
\ref{fig-6}. LAEs in our sample have a median dust-corrected $\tilde{SFR}(UV)=34$
M$_{\odot}$yr$^{-1}$, a factor of $\sim3$ higher than the uncorrected
median value, and show intrinsic $SFRs$ ranging
from 1 to 1500 M$_{\odot}$yr$^{-1}$.

The escape fraction of Ly$\alpha$
photons is given by the ratio between the Ly$\alpha$ derived $SFR$ and
the extinction corrected UV $SFR$.

\begin{eqnarray}
f_{esc}(Ly\alpha) &=& \frac{L(Ly\alpha)_{observed}}{L(Ly\alpha)_{intrinsic}} \nonumber \\
                  &=& \frac{SFR(Ly\alpha)}{SFR(UV)\times10^{(0.4k_{1500}E(B-V))}} \nonumber \\
\end{eqnarray}

\begin{figure*}[t]
\begin{center}
\plotone{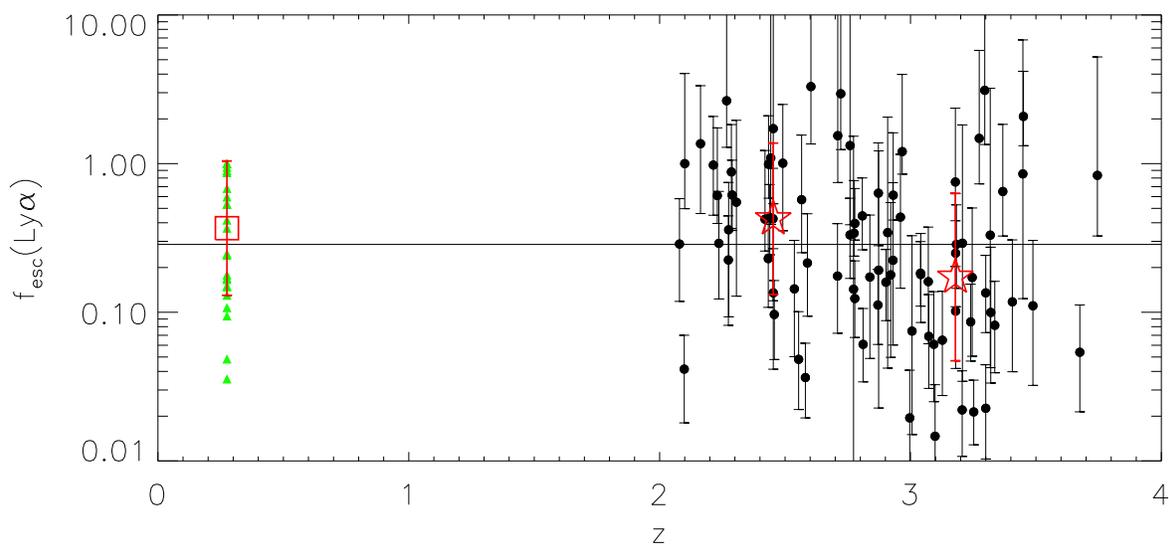}
\caption{Escape fraction of Ly$\alpha$ photons as a function of
  redshift for the 83 LAEs in the final sample. The solid horizontal line
  denotes the median escape fraction of 29\%. Also shown is the median
  escape fraction for the two redshift bins at $1.9<z<2.8$ and
  $2.8<z<3.8$ (open red stars), with error-bars corresponding to the
  standard deviation of log($f_{esc}$) within each bin. The escape fractions
  of LAEs at $z=0.3$ with their median from \cite{atek09} (green
  triangles, red open square) are also displayed.}

\label{fig-10}
\end{center}
\end{figure*}

Measured values of $f_{esc}(Ly\alpha)$ are reported in Table
\ref{tbl-1}. Figure \ref{fig-10} presents the Ly$\alpha$ escape
fraction of our LAEs
as a function of redshift (black circles). A broad range in the escape
fraction (2\% to 100\%) is observed. LAEs in our
sample show a median escape fraction
$\tilde{f}_{esc}(Ly\alpha)=0.29\pm0.04$,
and a mean escape fraction $\langle f_{esc}(Ly\alpha)\rangle =0.55\pm0.08$
(formal error on the mean). All objects showing $f_{esc}(Ly\alpha)>1$
are consistent with $f_{esc}(Ly\alpha)=1$ to 1.5$\sigma$.

A recent study by \cite{hayes10a} used a pair of optical and NIR
narrow-band filters to sample the Ly$\alpha$ and H$\alpha$ lines over
the same volume. By comparing the Ly$\alpha$ and H$\alpha$
luminosities of a sample of 38 LAEs at $z=2.2$, they derived a lower
limit of 0.32 for the average Ly$\alpha$ escape fraction of LAEs,
a value consistent with our measured average.
Another estimation of the Ly$\alpha$ escape fraction was done by
\cite{verhamme08} using an independent method on their
spectroscopic sample of 11 high-$z$ galaxies (8 of which are
LAEs). Fitting the Ly$\alpha$ emission line velocity profiles using Monte Carlo
radiative transfer simulations yielded best-fit values for
$f_{esc}(Ly\alpha)$ between 0.02 and 1, with a median value of 0.17,
in good agreement with our observed median value. The agreement between
these three independent estimations using a different set of
techniques is encouraging. 

A median (mean) escape fraction of $\sim20$\% ($\sim50$\%) is one order of magnitude higher than
that adopted in the semi-analytical models of \cite{ledelliou05}, in
which a 2\% escape fraction combined with a top-heavy IMF is
used to match the output of the models to the observed Ly$\alpha$ and
UV luminosity function of LAEs at different redshifts. We observe a
much larger escape fraction, and our measured $EW$s can be explained by standard
stellar populations with normal IMFs. Also, the large scatter seen in
Figure \ref{fig-10} implies that using a single value of
$f_{esc}(Ly\alpha)$ to model the LAE galaxy population is not a
realistic approach.

It is important to remark that estimating the escape fraction
directly from the observed $SFR(Ly\alpha)/SFR(UV)$ ratio, by assuming
LAEs are dust-free galaxies, would imply a significant
overestimation of its value. For example, the best-fit SED to the
stacked optical photometry of $z=3.1$ LAEs in \cite{gawiser06a},
which has $A_V=0^{+0.1}_{-0.0}$, implies a best-fit escape fraction of
0.8 (although the uncertainty in the fit allows for a escape fraction
$>0.2$, in agreement with our results). Similarly, the ratios measured by
\cite{ouchi08}, \cite{nilsson09a}, and \cite{guaita10} imply
escape fractions in the 0.5 to 1.0 range if dust is not
considered. As discussed above,
if we were to completely neglect dust extinction we, would measure a
value of 0.87 for our sample.

\subsection{Evolution of the Ly$\alpha$ Escape Fraction in LAEs}

No significant evolutionary trend is present across the $1.9<z<3.8$
range of our objects in Figure \ref{fig-10}, where the median escape
fractions for the two $1.9<z<2.8$ and $2.8<z<3.8$ redshift bins (red
open stars in Figure \ref{fig-10}) are
consistent with the median for the whole sample. In order to investigate if
the Ly$\alpha$ escape fraction of LAEs evolves over a larger baseline
in cosmic time, we also show results found in the literature at a lower
redshift. At higher redshifts the Ly$\alpha$ escape fraction for LAEs
remains poorly constrained, although attempts to measure it exist in
the literature \citep[eg.][]{ono10}
 
At low redshift \cite{atek09}
performed optical spectroscopy on a sample of $z\simeq 0.3$ LAEs \citep{deharveng08}
and used the H$\alpha$ luminosity, in combination with dust extinctions
derived from the Balmer decrement, to estimate the Ly$\alpha$ escape
fraction of these objects. A similar range in the escape fraction
is observed for $z\simeq0.3$ and $2<z<4$ LAEs, with the former showing
values ranging from 0.03 to 1, implying that there has not been significant
evolution in $f_{esc}(Ly\alpha)$ over the $\sim$8 Gyr from $z\sim3$ to
$z\sim0.3$. At very high redshifts
($z=5.7$ and 6.6) \cite{ono10} has estimated the Ly$\alpha$ escape
fraction of a sample of a few hundred narrow-band selected LAEs using
a similar method to the one used here, except
that their intrinsic $SFRs$ were measured by SED fitting. Their escape
fractions are consistent with our measured values at $2<z<4$,
although their error-bars are large. Therefore we detect no significant 
evolution in $f_{esc}(Ly\alpha)$ over the $0.3<z<6.6$ range. 

This lack of evolution in the Ly$\alpha$ escape fraction of LAEs must be
interpreted with caution, since nothing ensures that the LAE
selection technique recovers the same galaxy populations at these very
distant epochs in the universe. Furthermore, since the selection is
based on the strength of the Ly$\alpha$ line relative to the the
underlying continuum (i.e. the $EW$ of the line), the technique will
tend to favor galaxies with high Ly$\alpha$ escape
fractions, as long as they satisfy the brightness cut of the survey, at any
redshift. Therefore, the lack of evolution in $f_{esc}(Ly\alpha)$ cannot be
interpreted as constancy in the physical conditions in the ISM of
these galaxies. For example, while at low redshift the escape fraction
is most likely dominated by dust absorption, at $z\sim6$ it is most
likely dominated by IGM attenuation.

\subsection{The Relation between $f_{esc}$(Ly$\alpha$) and Dust}

As discussed in \S1, a major subject of debate regarding the escape of
Ly$\alpha$ photons from star-forming galaxies is the role played by
dust. It is not clear whether the resonant nature of the transition 
produces Ly$\alpha$ photons to be extincted more, less, or in the same
amount as continuum photons outside the resonance. For example, while
Ly$\alpha$ photons should originate in the same regions as H$\alpha$
photons, we have no reason to expect the extinction seen by Ly$\alpha$ photons to 
follow the nebular extinction relation
$E(B-V)_{stars}=0.44E(B-V)_{gas}$ seen for non-resonant hydrogen
transitions in star-forming galaxies at $z=0$ \citep{calzetti97}, since
resonant scatter makes the optical paths seen by Ly$\alpha$ completely
different from the one seen by other lines like H$\alpha$ or
H$\beta$. Furthermore, it has not been established if the above
relation holds at high redshift or not. 

In order to test this issue we parameterize the ratio between the dust
opacity seen by Ly$\alpha$ and that which continuum photons at the same
wavelength would see in the absence of the transition. Following
\cite{finkelstein08}, we adopt the parameter
$q=\tau_{Ly\alpha}/\tau_{\lambda=1216}$, where
$\tau_{\lambda}=k_{\lambda}E(B-V)/1.086$ and $k_{\lambda}$ is assumed
to be a \cite{calzetti00} dust attenuation law. $E(B-V)$ is always
taken to be the stellar color excess derived from the UV slope.

A value of $q\sim0$ implies that Ly$\alpha$ photons suffer very little
extinction by dust, as is expected in an extremely clumpy multi-phase ISM. Large
values ($q\gg 1$) represent cases in which scattering of Ly$\alpha$
photons introduces a strong increase in the dust attenuation as is expected
in a more homogeneous ISM. As discussed in \S1, not only the
structure of the ISM determines the value of $q$, but also the
kinematics of the ISM, since favorable configurations (eg. an expanding
shell of neutral gas which allows backscattering) can reduce
the amount of dust extinction seen by Ly$\alpha$ photons. 
All these processes are coupled in
determining the value of $q$, and discriminating between them
requires a joint analysis of the UV, Ly$\alpha$, and H$\alpha$
luminosities, the dust extinction either from the shape of the
UV continuum or from Balmer decrements, and the profiles of the latter
two emission lines. Until such data exist, interpretation is
difficult, but we can still gain insights about the escape of Ly$\alpha$ photons
and the dust properties of LAEs from the measured value of $q$.
 
\begin{figure*}[t]
\begin{center}
\plotone{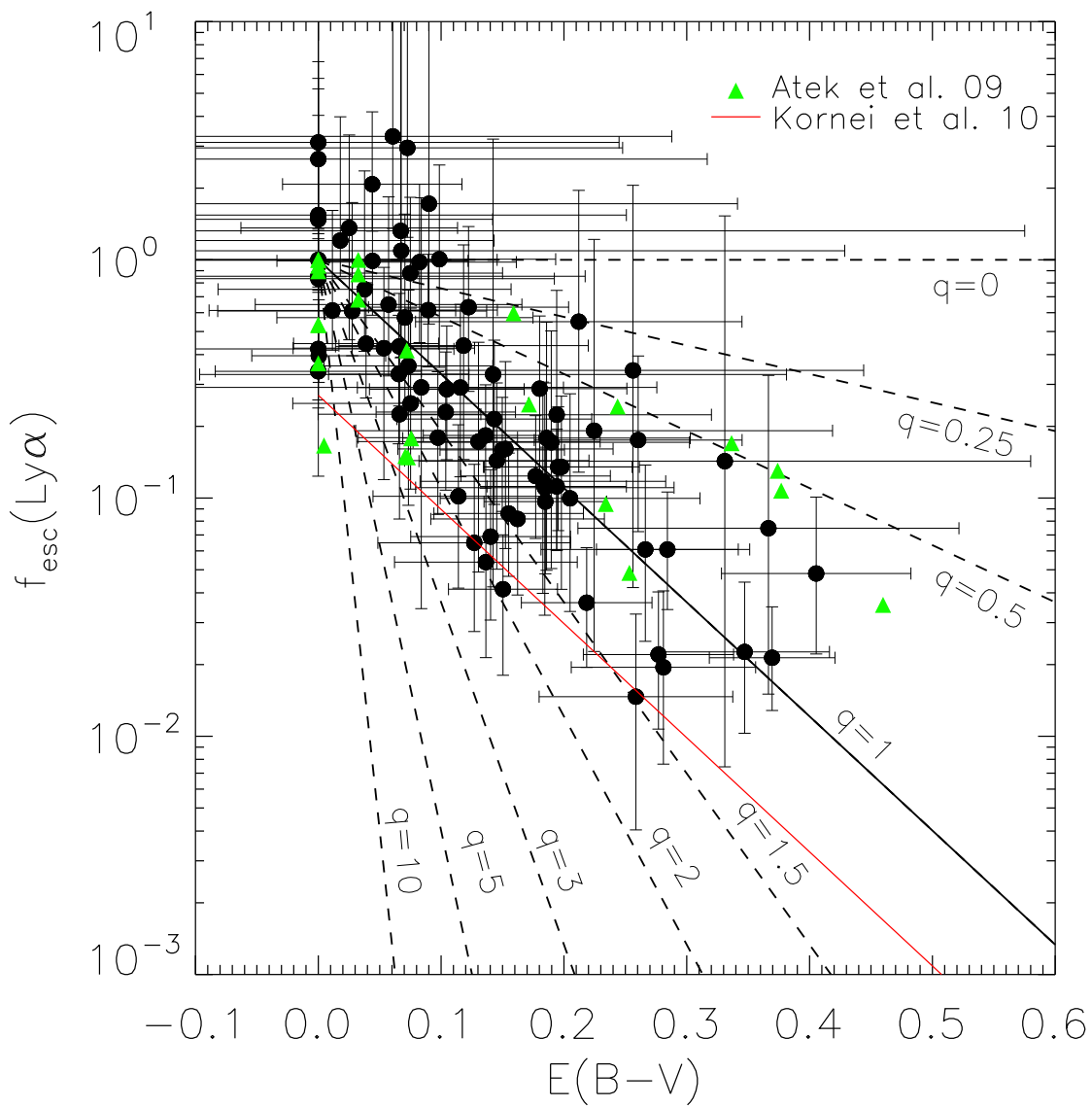}
\caption{Ly$\alpha$ escape fraction as a function of E(B-V). Dashed
  lines show the expected correlation for different values of the
  parameter $q=\tau_{Ly\alpha}/\tau_{\lambda=1216}$. The red line
  displays the relation for LBGs showing Ly$\alpha$ in emmission
  from \cite{kornei10}. Green triangles show the values for $z\simeq0.3$ LAEs from \cite{atek09}.}
\label{fig-11}
\end{center}
\end{figure*}

In Figure \ref{fig-11} we present the Ly$\alpha$ escape fraction versus
the dust reddening $E(B-V)$ for our sample of LAEs. A clear
correlation between the escape fraction and the amount of dust
extinction is seen. Also shown are the
expected correlations for different values of the parameter $q$. The
LAE population falls along
the $q=1$ relation. The median for the whole sample is
$\tilde{q}=0.99\pm0.44$ (median
absolute deviation error), implying that in most LAEs the Ly$\alpha$
emission suffers a very similar amount of dust extinction to that
experienced by UV continuum light. 

Our results show good agreement with those of \cite{hayes10a}. In their
work, 5 out of 6 LAEs at $z=2.2$ detected in both Ly$\alpha$ and H$\alpha$
show escape fractions and dust reddenings consistent with the $q=1$
relation. The same is true for the large majority of their LAEs with
no H$\alpha$ detections for which they could only provide lower limits
to the escape fraction. Our LAEs at $2<z<4$ also occupy the same
region in the $E(B-V)$ vs. $f_{esc}(Ly\alpha)$ plane as the low redshift
LAEs \citep[$z\simeq0.3$][]{atek09} shown as green triangles in Figure
\ref{fig-11}. This implies that not only the Ly$\alpha$ escape
fraction in LAEs does not evolve with redshift as shown in \S5.3, but
its dependence on the dust content of the ISM remains the same from
$z=4$ to 0.3. 

The LAE selection criteria imply that these objects are chosen to be
the galaxies with the largest Ly$\alpha$ escape fractions given their
Ly$\alpha$ luminosities and dust content at any
redshift. Most likely, a combination of ISM geometry and kinematics favors the
escape of Ly$\alpha$ photons in these galaxies as compared to the
common galaxy population at any redshift. Hence, when examining
Figure \ref{fig-11} we should think of LAEs as the upper envelope of
the escape fraction distribution at any E(B-V). For example,
\cite{kornei10} found that LBGs with Ly$\alpha$ emission typically lie below the
$q=1$ relation. In their work they parameterized the difference in the
extinction seen by Ly$\alpha$ and continuum photons by the ``relative
escape fraction'', $f_{esc,rel}$, which relates to $q$ by the following
relation

\begin{equation}
q=\frac{log(f_{esc,rel})}{0.4k_{\lambda=1216}E(B-V)}
\end{equation}

\noindent
They find LBGs to have $\langle f_{esc,rel}
\rangle=0.27$ (which does not include LBGs showing Ly$\alpha$ in absorption). We
present this relation for LBGs as a red line in Figure
\ref{fig-11}. This finding supports our previous point, namely that LAEs are
the upper envelope to the overall galaxy population in
the $E(B-V)$ vs. $f_{esc}(Ly\alpha)$ plane. 

Our result should not be interpreted as evidence against the existence of a
clumpy multiphase ISM in LAEs, since in the presence of a completely
homogeneous ISM we expect $q>1$. Nevertheless, our result indicates that either
a clumpy ISM, a favorable kinematic configuration of the ISM, or a
combination of both, can reduce the amount of dust seen by
scattering Ly$\alpha$ photons only up to the point where they encounter the
same level of dust opacity as the continuum. Since LAEs by definition
will be the galaxies with the largest Ly$\alpha$ escape fractions at
any value of $E(B-V)$, the absence of a significant number of points
at low values of $q$ in Figure \ref{fig-11} suggests that enhancement of the
Ly$\alpha$ $EW$s due to clumpiness in the ISM is not a common
process in galaxies.

\section{The L\lowercase{y}$\alpha$ Luminosity Function}

It has been well established in the literature that the Ly$\alpha$
luminosity function does not show significant evolution from
$z=3$ to 6 \citep{shimasaku06, tapken06, ouchi08}. On the other hand,
a strong decrease of roughly one order of magnitude is seen in the
abundance of LAEs from $z\sim 3$ down to $z\sim0.3$
\citep{deharveng08}. At what point in cosmic time 
this decrease starts to take place, and how well it traces the
$SFR$ history of the universe, is unknown. Recently there have
been reports of possible evolution in the number density of LAEs
between $z\sim3$ and $z\sim2$ \citep{nilsson09a}, but, as
stated by the same authors, these results might be affected by cosmic
sample variance over the surveyed volumes. Furthermore, \cite{cassata11}
find no evolution in the luminosity function across these epochs in their sample of
spectroscopically detected LAEs. The existence of evolution
in the luminosity function (or number density) of LAEs down to these
lower redshifts is still a subject of debate. 

By examining the redshift distribution of
our LAEs (Figure \ref{fig-2}), we found indications that their
number density might be decreasing towards lower redshifts ($z<3$) in our
sample (\S3). In this section we measure the Ly$\alpha$ luminosity function of LAEs,
and study any potential evolution down to $z\sim2$. We
restrict the measurement of the luminosity function to LAEs in the COSMOS and GOODS-N
fields, which account for 81\% (80/99) of the total sample. Both
MUNICS and XMM-LSS lack deep X-ray data comparable to the one
available in COSMOS and GOODS-N, so it is not possible to identify
AGNs in those fields.

\subsection{Measurement of the Luminosity Function}

To measure the luminosity function we adopt a $1/V_{max}$ formalism
similar to the
one used by \cite{cassata11}. We compute the
volume density of objects in bins of $\Delta log(L)=0.125$ dex, as the sum
of the inverse of the maximum volumes over which each object in the
luminosity bin could have been detected in our survey. As discussed in
\S3, the depth of the observations is variable across the
surveyed area. The whole survey covers 169 arcmin$^2$, corresponding to 60
VIRUS-P pointings. Each pointing was covered by six dithered
observations, which accounts for 360 independent observations each
reaching different depths. The noise spectrum for each IFU fiber in
each of these observations is an output of our data processing
pipeline, and can be directly translated into an effective line luminosity
limit for Ly$\alpha$ at each redshift (see Figure \ref{fig-1}).

\begin{figure*}[t]
\begin{center}
\plotone{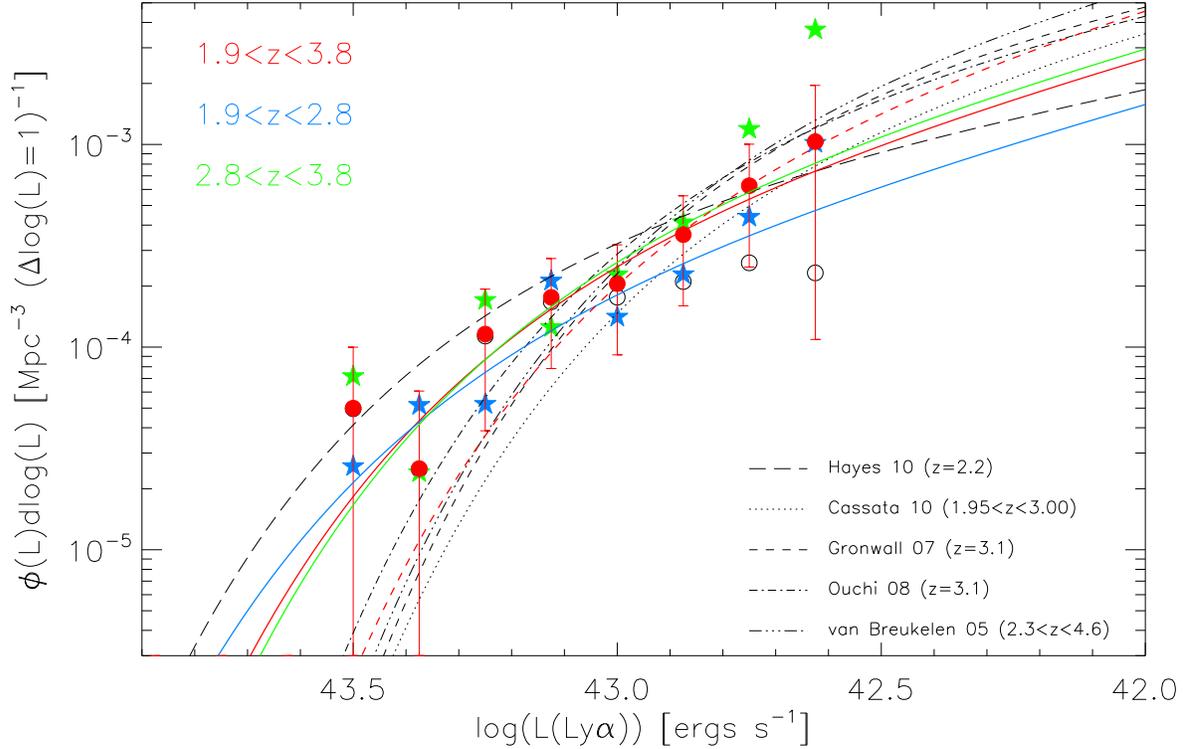}
\caption{Ly$\alpha$ luminosity function of the HETDEX Pilot
  Survey sample of 80 LAEs in COSMOS and HDF-N, shown before and after applying the
  completeness correction (open black and filled red circles
  respectively). Poisson error-bars are included. Also displayed are the
  completeness corrected luminosity function for the two redshift bins
  at $1.9<z<2.8$ and $2.8<z<3.8$ (blue and green stars respectively),
  and the luminosity functions of \cite{vanbreukelen05, gronwall07, ouchi08,
  hayes10a},and \cite{cassata11}. Schechter fits to the full sample, as
  well as the low-z and high-z samples, are also presented (solid red,
  blue, and green curves respectively). The red dashed line
  denotes the best Schechter fit to the $L(Ly\alpha)\leq10^{43}$erg
  s$^{-1}$ bins.}
\label{fig-12}
\end{center}
\end{figure*}

For each object, $V_{max}$ is given by

\begin{equation}
V_{max}=\displaystyle \sum_iV_{max,i}
\end{equation}

\noindent
where $V_{max,i}$ is the integral of the co-moving volume over all the
redshifts at which the object could have been detected (i.e. where the
luminosity limit is lower than the objects luminosity) for each
observation $i$. Summing over the inverse of $V_{max}$ for all
objects in each luminosity bin then yields the luminosity function
shown as open black circles in Figure \ref{fig-12}.

As mentioned briefly in \S3 and discussed extensively in Paper I,
the effects of incompleteness are important over all luminosities in
our survey. Completeness is a direct function of the S/N at which the
emission line is detected in our spectra. 
Since the limiting luminosity is not constant at
all redshifts (Figure \ref{fig-1}), objects of the same
luminosity can be detected with high significance, and hence high
completeness, at certain redshifts and with low significance and
low completeness at others. This is different than, for example,
imaging surveys where objects are detected in a narrow redshift range,
and the S/N is close to a unique function of the luminosity. In those cases,
incompleteness becomes only important in low luminosity bins, where the
objects flux approach the depth of the images\footnote{In
  reality, incompleteness in narrow-band emission line surveys is more complicated
  than this because of the non top-hat shape of narrow-band filters,
  and shows a dependence with the redshift of the sources; see the
  discussion in \cite{gronwall07}.}. In our case, we must
account for incompleteness over the whole luminosity range if we want
to get a proper estimate of the luminosity function.

In Paper I we present a detailed completeness
analysis of our survey based on simulations of the recovery
fraction of synthetic emission lines at different S/N in our
spectra. Using these recovery fractions we correct the observed
Ly$\alpha$ luminosity function calculated as described above. The resulting
completeness-corrected luminosity function is shown by the red filled circles in
Figure \ref{fig-12}. Error-bars shows Poisson uncertainties only, and
correspond to a lower limit for the error since they do not include
cosmic variance, although \cite{ouchi08} show that for volumes such
as the one surveyed here ($\sim 10^6$ Mpc$^3$), cosmic sample variance
uncertainties are not significantly higher than Poisson errors.

We fit the observed Ly$\alpha$ luminosity function using a \cite{schechter76} function
of the following form

\begin{equation}
\phi(L)dL=\phi^*(L/L^*)^{\alpha}\;exp(-L/L^*)\;d(L/L^*)
\end{equation}

Since the depth of our observations ($\sim 5\times10^{-17}$ erg
s$^{-1}$cm$^{-2}$ in line flux) is somewhat limited, we do not consider
our data to be sufficiently deep to constrain the faint-end slope
($\alpha$) of the luminosity function. We consider the best available constraint on
$\alpha$ to come from the spectroscopic survey recently performed by
\cite{cassata11}. They measure $\alpha\simeq-1.7$ using a survey which
reaches more than one order of magnitude deeper than ours in terms of
limiting line flux ($\sim 1.5\times10^{-18}$ erg
s$^{-1}$cm$^{-2}$). We take their measured $\alpha$ as our
fixed fiducial value for the faint-end slope of the luminosity function, but also report results
assuming $\alpha=-1.5$, since that is the value most commonly used in
the literature \citep{gronwall07, ouchi08}. Our results do not depend
significantly on the assumed value of $\alpha$.

The best-fit Schechter luminosity function ($\alpha=-1.7$) is shown by the solid red line
in Figure \ref{fig-12}, and 1, 2, and 3$\sigma$ confidence limits for
the parameters are shown in Figure \ref{fig-13}. The best fit
parameters for $\alpha=-1.7$ and -1.5 are reported in the first two
rows of Table \ref{tbl-2}.

\subsection{Comparison with Previous Measurements}

Figure \ref{fig-12} also shows the Ly$\alpha$ luminosity functions measured by several authors at
similar redshifts. The overall agreement with previous measurements is
good. The Ly$\alpha$ luminosity functions of \cite{vanbreukelen05, gronwall07,
  ouchi08, hayes10a}, and \cite{cassata11} measured at $2.3<z<4.6$,
$z=3.1$, $z=3.1$, $z=2.2$, and $1.95<z<3$ respectively, agree with our observed values to
within $\sim 1\sigma$ (Poisson) at all luminosities. The
\cite{hayes10a} measurement shows better agreement with our data at
the bright end of the luminosity function. This is in fact surprising,
as their measurement was performed over a smaller volume
($5.4\times10^3$ Mpc$^3$) and a fainter range in luminosities
($2\times10^{41}-5\times10^{42}$ erg s$^{-1}$) than
the other mentioned works.

\begin{figure}[b]
\begin{center}
\plotone{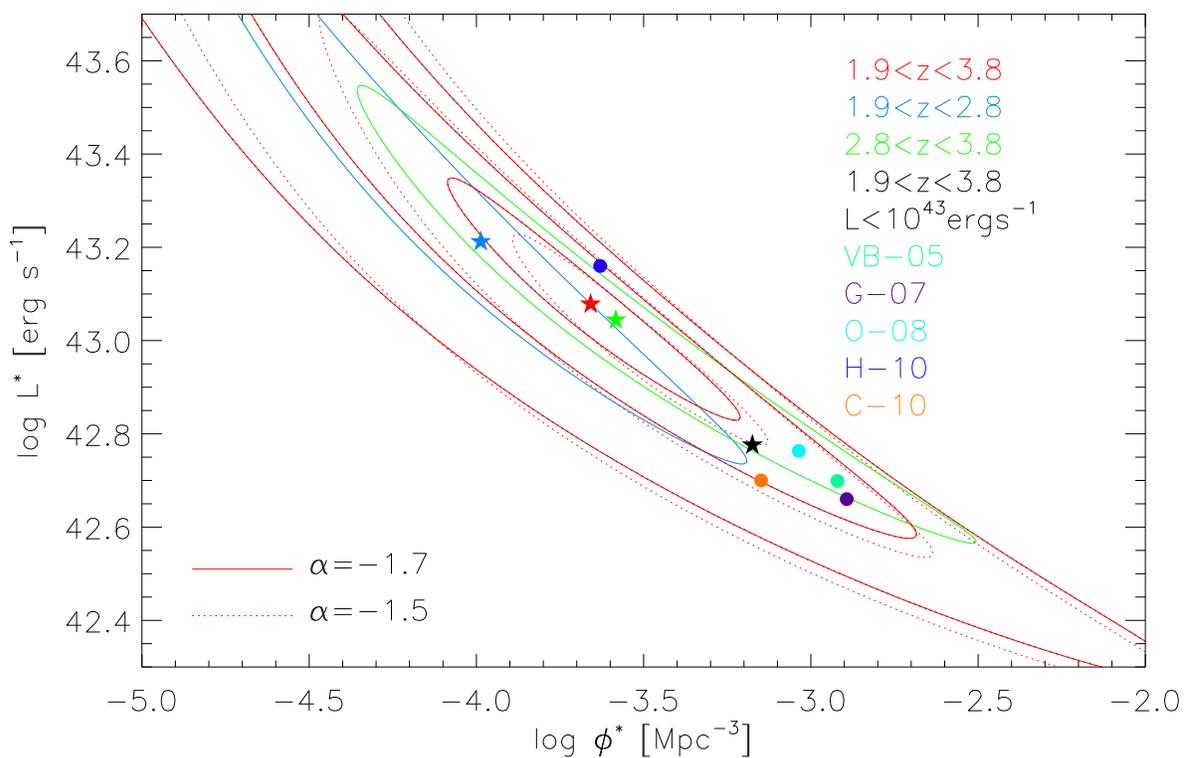}
\caption{Contours show 1, 2, and 3$\sigma$ confidence limits for the
  luminosity function parameters $L^*$ and $\phi^*$. Stars show our results for the full
  sample and the two redshift bins
  at $1.9<z<2.8$ and $2.8<z<3.8$. The
  parameters estimated by \cite{vanbreukelen05, gronwall07, ouchi08,
  hayes10a}, and \cite{cassata11} are also presented (filled circles).}
\label{fig-13}
\end{center}
\end{figure}

Our best-fit Schechter function appears to be flatter than previous
measurements over a similar range in luminosities.  Figure
\ref{fig-13} shows that this difference is because we derive a 
higher $L^*$ and a lower $\phi^*$ than previous
authors (except \cite{hayes10a} who found a very similar value for
$\phi^*$ but a larger value for $L^*$). The best-fit
parameters measured by \cite{vanbreukelen05, gronwall07, ouchi08, hayes10a}, and 
\cite{cassata11} fall within our 2$\sigma$ confidence contour. This
last work is the only one of the
mentioned luminosity function measurements in which $\alpha=-1.7$. For all the other
measurements, the faint-end slope was either assumed or measured to be $-1.5$ except for
\cite{vanbreukelen05} who used $-1.6$. For better comparison, Figure \ref{fig-13} also shows
uncertainty contours for our fit assuming $\alpha=-1.5$ (dotted contours). As mentioned
above, the value of $\alpha$ does not change our results in any
significant way.

The $L^*$ and $\phi^*$ parameters are strongly correlated with each other, so the
$2\sigma$ discrepancy with previous measurements is not surprising as it follows the sense of the
correlation. Most importantly, we survey a very large volume and hence
are able to find rare high luminosity objects. The luminosity
functions derived in these studies typically stop at
$\sim10^{43}$ erg s$^{-1}$, while we find objects up to three times brighter
luminosities. If we fit a Schechter function to only bins with $L(Ly\alpha)
\leq 10^{43}$ erg s$^{-1}$, we obtain the luminosity function shown as a
dashed red line in Figure \ref{fig-12}, which is in much better agreement
with previous measurements (black star in Figure \ref{fig-13} and
third row in Table \ref{tbl-2}).

\subsection{Evolution of the Ly$\alpha$ Luminosity Function}

As mentioned above, evidence suggests that the Ly$\alpha$ luminosity
function does not significantly evolve between $z\sim3$ and
$z\sim6$. While at the high end of this redshift range ($z\gtrsim$5)
IGM absorption might become important and the lack of evolution might
imply an increase in the intrinsic Ly$\alpha$ luminosity function
\citep{cassata11}, at least between $z\sim4$ and $z\sim3$ the lack of
intrinsic evolution seems well supported as changes in IGM transmission
are negligible \citep{ouchi08}. We can extend these studies to lower
redshifts and ask: Does the luminosity function show any significant
evolution down to $z\sim2$?

To test for possible evolution, we have again divided our sample in the two
redshift bins defined in \S4, one at low-redshift ($1.9<z<2.8$), and
another one at high-redshift ($2.8<z<3.8$). We measure the luminosity
function in each of these bins independently.
The results are shown in Figure \ref{fig-12}, best fit parameters
are presented in Table \ref{tbl-2}, and 1$\sigma$ confidence limits
are shown in Figure \ref{fig-13}. At $L(Ly\alpha)\leq 10^{43}$ erg s$^{-1}$,
where cosmic variance is lower than at higher luminosities, the low-$z$
luminosity function seems to be systematically lower
than the high-$z$ luminosity function by a factor about $\sim2$, in
agreement with what we observed in \S3 when comparing the redshift
distribution of our objects to the predictions for a non evolving
luminosity function. Still, both the high-$z$ and low-$z$ luminosity
functions fall within their mutual Poisson uncertainties, and
there is overlap between the 1$\sigma$ confidence limits in their best-fit
Schechter parameters (Figure \ref{fig-13}).

We conclude that we find indications for evolution in the
luminosity function over the $2\lesssim z\lesssim4$ range, with a
decrease towards lower redshifts, but only at a low significance
level. Larger samples, such as the ones HETDEX will produce in its few
firsts months of operation, will be required to confirm this. If real,
this evolution implies that the large drop in the Ly$\alpha$
luminosity function, evident at $z\simeq0.3$, starts to occur at
$z>2$. Another way of searching for evolution in the Ly$\alpha$ luminosity
function is to integrate it, and compare the implied Ly$\alpha$
luminosity density at each redshift. This is the subject of the next section.

\section{Evolution of the L\lowercase{y}$\alpha$ Luminosity Density and
  the Global Escape Fraction of Ly$\alpha$ Photons.}

In \S5.2 we measured the median escape fraction of Ly$\alpha$ photons
at $2<z<4$ in LAEs to be $\sim30$\%. This does not
represent the median escape fraction of the whole galaxy population at
those redshifts, since LAEs will, by definition, be biased towards
having high $f_{esc}(Ly\alpha)$. On the other hand, we can integrate
the Ly$\alpha$ luminosity function measured in the previous section to
estimate the Ly$\alpha$ luminosity density ($\rho_{Ly\alpha}$) at these
redshifts. Comparing this observed luminosity density with that
predicted from the global $SFR$ density ($\rho_{SFR}$) for the entire
galaxy population provides an estimate of the global escape
fraction of Ly$\alpha$ photons and its evolution with redshift. 
The above approach is equivalent to taking the ratio between the $SFR$
density implied by the observed Ly$\alpha$ luminosity density using
Equation 2 ($\rho_{SFR,Ly\alpha}$), and the total $\rho_{SFR}$. This
method has been applied by \cite{cassata11}. In this work we extend
their analysis which included the \cite{cassata11} data at $2<z<6.6$,
the measurement of \cite{gronwall07} at $z=3.1$, and the data of
\cite{ouchi08} at $z=3.1$, 3.7, and 5.7. We add the HETDEX Pilot
Survey data points at $1.9<z<3.8$, as well as the $z\sim0.3$ LAE data from
\cite{deharveng08} and \cite{cowie10}, the $z=2.2$ data of
\cite{hayes10a}, the $z=4.5$ measurement by \cite{dawson07}, the
measurement at $z=5.7$ of \cite{shimasaku06}, the $z=6.5$ data from
\cite{kashikawa06}, the data of \cite{ouchi10} at z=6.6, and the
$z=7.7$ measurement of \cite{hibon10}. A similar dataset has been analyzed
in this way in a recent submission by \cite{hayes10b}, although using
a different set of H$\alpha$ and UV luminosity functions at different
redshifts to estimate the total SFR density.

The top panel in Figure \ref{fig-14} shows $\rho_{SFR,Ly\alpha}$ derived from the observed
Ly$\alpha$ luminosity density using Equation 2. We present our results
for the full sample and for the low-$z$ and high-$z$ bins of the HETDEX
Pilot Survey (red, blue, and green filled circles), as well as the
compilation of data points calculated from the
Ly$\alpha$ luminosity functions at $0.3<z<7.7$ mentioned above (black
filled circles). Vertical error-bars are estimated from the
published uncertainties in $L^*$ and $\phi^*$, and horizontal error-bars
show the redshift range of the different samples (omitted for
narrow-band surveys). Also presented is the latest estimate of the total
$SFR$ density history of the universe from \cite{bouwens10b}, which has
been derived from the best to date collection of dust extinction corrected
UV luminosity functions at a series of redshifts between 0 and 8, and
shows a typical uncertainty of 0.17 dex \citep[][ and reference
  therein]{bouwens10b}.

A source of systematic error in measuring the Ly$\alpha$ luminosity
density comes from the choice of the luminosity limit down to which
the integration of the luminosity function is performed. An excellent
discussion on this issue can be found in \cite{hayes10b}. With the
goal of estimating the volumetric Ly$\alpha$ escape fraction by
comparison to the UV derived SFR density, we should ideally choose an
integration limit consistent with the one used by \cite{bouwens10b} to
integrate their UV luminosity functions. In this way we can ensure
both measurements are roughly tracing the same galaxies. In the case
of Ly$\alpha$ and UV luminosity functions this is nontrivial, as the
exact number will depend on the, mostly unconstrained, shape of the
Ly$\alpha$ escape fraction distribution for galaxies as a function UV
luminosity. In lack of a better choice, we follow the approach of
\cite{hayes10b}, and integrate the Ly$\alpha$ luminosity functions
down to the same fraction of $L^*$ as the UV luminosity function were
integrated (0.06$L_{z=3}^*$ in the case of \cite{bouwens10b}). For
consistency with \cite{hayes10b}, and in order to allow for a better
comparison with their results, we define this limit using the
\cite{gronwall07} luminosity function at $z=3.1$, for which the
integration limit becomes $0.06L_{G07}^*=2.66\times10^{41}$ erg s$^{-1}$. 
For all the data points in Figure \ref{fig-14} we also shows the same
measurements obtained by integrating the luminosity functions down to
zero luminosity (upper open circles). The unlimited integration
typically overestimates the luminosity (SFR) density by $\sim60$\%. 
This provides a notion of the maximum impact that the choice of this
integration limit has on the measured value of the luminosity density. 

A second source of systematic error in the above measurement comes from
the role that IGM scattering has at reducing the observed Ly$\alpha$
flux of sources at very high redshifts. Although all our previous analysis
neglected the effects of IGM scattering on the Ly$\alpha$ line, this
approach was only well justified at our redshifts of interest ($z<4$),
where IGM scattering is negligible for our purposes (see discussion in
\S1 and \S5). To study the escape of Ly$\alpha$ photons from galaxies
across a larger redshift range, we should try to incorporate the
effect of the IGM, which is not negligible for the measurements at very
high redshift ($z\sim6$). As discussed in \S1, the effects of the ISM
and IGM kinematics in and around galaxies makes this correction very
difficult \citep{dijkstra07, verhamme08}. To first order, we have
applied a correction using the \cite{madau95} average Ly$\alpha$
forest opacity, and assuming that only half of the Ly$\alpha$ line
flux suffers this attenuation. The filled symbols in Figure
\ref{fig-14} include this correction. Raw measurements done without
applying this correction are also shown in Figure \ref{fig-14} as the
open circles below each data point. While this correction can become large
($\sim50$\%) at the highest redshifts, it impact is still within the
1$\sigma$ uncertainties coming from the luminosity function measurements.

In accordance with the low significance hint of evolution presented in \S6.3, in the $2<z<4$ range, all
the estimates of $\rho_{SFR,Ly\alpha}$ agree with each other within 1$\sigma$. 
However, by examining the overall trend of the data points, and keeping in
mind the ones at higher and lower redshifts, there are clearly
indications for evolution in the $SFR$ density derived from Ly$\alpha$ from $z\sim7$
down to $z\sim0.3$, with a steady decrease towards lower redshifts across the $2<z<4$
range. Although the uncertainties in the $2<z<4$ range
are large, allowing any two datapoints to be consistent with each other, the
overall trend implies a decrease in $\rho_{Ly\alpha}$ of
roughly a factor of $\sim2$ from $z=4$ to 2. We stress the need for
larger samples of LAEs at these redshifts to better constrain this evolution.

The bottom panel of Figure \ref{fig-14} shows the global average escape
fraction of Ly$\alpha$ photons, which is given by the ratio between
$\rho_{SFR,Ly\alpha}$ and $\rho_{SFR}$ at any given redshift. The
average escape fraction derived from our Ly$\alpha$ luminosity function over
the whole $1.9<z<3.8$ range is ($3.0^{+2.3}_{-1.2}$)\%. Errors include
1$\sigma$ uncertainties in the luminosity function parameters and the
0.17 dex uncertainty in the total SFR density from
\cite{bouwens10b}. For our
$1.9<z<2.8$ and $2.8<z<3.8$ bins we derived a mean
Ly$\alpha$ escape fraction for the overall galaxy population of
$(2.0^{+2.0}_{-0.9})$\% and $(4.3^{+10.3}_{-2.2})$\% respectively. This
amount of evolution is not statistically significant, but we believe
it to be real in the context of the overall trend seen in Figure \ref{fig-14}.
It also does not contradict
the lack of evolution in the escape fraction for LAEs observed in
\S5.2, since, as mentioned above, the LAE selection tends to identify
galaxies with high $f_{esc}(Ly\alpha)$ at any redshift, independent of
the value of the escape fraction of the total galaxy population.
The median dust extinction of a factor of $\sim3$ measured
in \S4.2 implies that LAEs contribute roughly 10\% of the
total star formation at $2<z<4$. This contribution rises to
80\% by $z\sim6$, implying that galaxies at these redshifts
must have very low amounts of dust in their ISM, which is consistent
with the very blue UV slopes of continuum selected galaxies at
these high redshifts \citep{bouwens10a, finkelstein10}. The
observed behavior is also consistent with the results of
\cite{stark10}, who find the fraction of LBGs showing high $EW$
Ly$\alpha$ emission to roughly double from $z=4$ to 6. A similar result
was also reported by \cite{ouchi08}, who measure a significant level of
evolution in the UV luminosity function between their $z=3.7$
and $z=6.6$ samples of LAEs, which was not traced by the Ly$\alpha$ luminosity
function.

The above escape fractions are in agreement with the result of
\cite{hayes10a}, who measured an overall Ly$\alpha$ escape fraction of
$(5.3\pm3.8)$\% at $z=2.2$ by comparing the Ly$\alpha$ and H$\alpha$
luminosity function of narrow-band selected emission line galaxies
over the same co-moving volume. On the other hand, by applying the same
method used here \cite{cassata11} measured an average escape
fraction of $\sim$20\% at $z=2.5$. The difference is easily explained
by the fact that the latter authors compared their observed Ly$\alpha$
derived $SFR$ density (which agrees with our value) to the total $SFR$
density uncorrected by dust, which underestimates the true value at
these redshifts.

It is evident that a strong decrease
in the Ly$\alpha$ escape fraction of galaxies occurred between
$z\sim6$, and $z\sim2$. In order to quantify this decrease we fit the
data points in the lower panel of Figure \ref{fig-14} using two
different functional forms. First, we fit a power-law of the form 

\begin{equation}
{\rm log}(f_{esc}(z))={\rm log}(f_{esc}(0))+\xi\;{\rm log}(1+z) 
\end{equation}

This is the same parametrization used by \cite{hayes10b} to fit the
history of the global Ly$\alpha$ escape fraction. Best fit parameters
are presented in Table \ref{tbl-3}. In order to provide a quantitative
assessment of the impact of systematic errors in the measurement, we
not only fit our best estimates of the escape fraction at each
redshift, but also the values calculated ignoring the luminosity
function integration limit, and the IGM correction. The best fitting
power-laws for these three sets of data points are shown as dotted
lines in Figure \ref{fig-14}. For comparison with the results of
\cite{hayes10b}, we should consider our raw measurement without
including the effects of the IGM, as a correction of this type was not
done in their work. They find best fit values of ${\rm
  log}(f_{esc}(0))=-2.8\pm0.1$, and $\xi=2.6\pm0.2$, in excellent
agreement with our result.

While the power-law model provides a reasonable fit to the
data, it seems to systematically overestimate the measured values of
$f_{esc}(Ly\alpha)$ in the $2<z<5$ range, and underestimate them in
the $5<z<8$ range. The data points in Figure \ref{fig-14} seem to
indicate a sudden drop, or transition in the escape fraction between 
$z=6$ and 2. A similar transition, in a coincident redshift range, is
observed in the dust extinction derived from the UV slope of continuum
selected galaxies \citep{bouwens09}. Given the important role that dust
has at regulating the escape of Ly$\alpha$ photons, it would not be
surprising if the dust content and the Ly$\alpha$ escape fraction of
galaxies present a similar evolution with redshift. In order to
quantify this behavior we also fit a {\it transition} model of the
following form,

\begin{equation}
{\rm log}(f_{esc}(z))=\frac{{\rm log}(f_{esc}(0))}{2}\;(1-{\rm tanh}(\theta(z-z_{tr}))
\end{equation}

\noindent
where $z_{tr}$ is the transition redshift at which the decrease in the escape
fraction takes place ($f_{esc}=1$ for $z\gg z_{tr}$), and the
parameter $\theta$ determines the sharpness of the
transition. Best-fit parameters to the measured escape fraction at
each redshift, and the values without IGM correction, and without a
luminosity function integration limit are presented in Table \ref{tbl-3}. 
Our best-fit {\it transition} model, implies a very high
Ly$\alpha$ escape fraction of $\sim80$\% at $z\sim6$, which drops
softly from $z\sim6$ to $z\sim2$, with a characteristic transition
redshift at $z_{tr}=4.0\pm0.5$, in order to reach a value of $\sim1$\%
in the local universe.

By analyzing the values in Table 3, It can
be seen that, given the current uncertainties, the IGM correction and
the choice of the luminosity function integration limit do not induce
major changes in the best-fit parameters, especially in the case of
the power-law exponent. The largest effect is that of the integration
limit on the escape fraction at $z=0$. The reason for this is that low
$L^*$ values are measured for the Ly$\alpha$ luminosity functions at
low redshift. Therefore, the integration limit lays closer to $L^*$ at
these redshifts, making the value of the luminosity density more
dependent on it.

Equation 8 predicts the average Ly$\alpha$ luminosity of star-forming
galaxies at any redshift given their average $SFR$, and it might prove
useful for semi-analytic
models of galaxy formation attempting to reproduce the Ly$\alpha$
luminosity function. However, the escape fraction shows a very large
scatter for single objects, and it might be systematically different
for galaxy populations selected using different methods. Therefore, this
relation should be used with caution, and only in an statistical
manner. Also, this equation is only valid over the redshift range in
which observations are available, and to the extent that current
uncertainties allow. For example, given the current uncertainties, we
do not consider the escape fraction to be properly constrained at
$z>6.6$. While it is tempting to interpret the slight drop seen in
the last data point at $z=7.7$ as a possible reduction in
Ly$\alpha$ transmission due to the neutralization of the IGM as we
walk into the end of re-ionization, the error-bars are too large to
allow for any significant conclusions.

\section{Conclusions}

For a sample of LAEs at $1.9<z<3.8$,
detected by means of blind integral field spectroscopy of blank
extragalactic fields having deep broad-band optical imaging, we were
able to measure the basic quantities $SFR$, $E(B-V)$, UV luminosity,
Ly$\alpha$ $EW$, and $f_{esc}(Ly\alpha)$. From these quantities and
the correlations observed between them we conclude:

\begin{itemize}

\item Over the $2<z<4$ range LAEs show no evolution in the average
  dust content of their ISM, parameterized by the dust reddening
  $E(B-V)$, and measured from the UV continuum slope. These objects
  show a mean $\langle E(B-V) \rangle=0.13\pm0.01$, implying that dust
  absorbs $\sim70$\% of the UV photons produced in these
  galaxies. While one third of LAEs down to our luminosity limit are
  consistent with being dust-free, the level of dust extinction
  measured for the rest of the sample is significant, and should not
  be neglected.

\item At $z<3$, we see the possible appearance of a rare (6/89
  objects) population of highly reddened ($E(B-V)>0.45$) LAEs,
  typically showing high Ly$\alpha$ $EW$s. Two of
  these objects are in the MUNICS field where we do not have deep
  X-ray data to reject AGNs from our sample. The remaining four
  objects show low association probabilities ($\leq0.6$) with their
  broad-band counterparts, casting doubt on the validity of our UV
  slope  and $EW$ measurements. The presence of these objects in the sample does not affect
  significantly the average dust properties of LAEs at the low
  redshift end of our range. If real, these objects are of great
  interest since their presence could indicate that the fraction of
  dusty LAEs grows towards lower redshifts. Followup of these
  objects is necessary to confirm this.

\item The Ly$\alpha$ $EW$s of LAEs in our sample are consistent with the
  expectations for normal stellar populations with metallicities
  within 1/50 $Z_{\odot}$ and solar. We do not find it necessary to
  invoke a top-heavy IMF, the presence of
  population III stars, or an enhancement of the $EW$ due to a
  clumpy dust distribution in a multi-phase ISM, to explain the observed
  $EW$s.

\item LAEs in our sample show a median uncorrected UV derived
  $SFR\simeq 11$ M$_{\odot}$yr$^{-1}$. Correcting the UV luminosities
  for dust extinction increases this median value to  $SFR\simeq 34$
  M$_{\odot}$yr$^{-1}$, implying that assuming LAEs to be dust-free
  galaxies can translate into large underestimates of their
  $SFR$s. The ratio between the observed (i.e. uncorrected for dust)
  UV and Ly$\alpha$ derived
  $SFRs$ shows a median value of 0.83. Neither this ratio, nor the
  Ly$\alpha$ $EW$, show significant evolution with redshift across the
  $2<z<4$ range. These two quantities also show no dependence with
  $E(B-V)$, implying
  that the ratio between the amount of dust extinction seen by
  Ly$\alpha$ photons and that seen by UV photons is independent of the
  dust-content of the galaxies' ISM. This finding is at odds with the
  expectation of models in which a clumpy distribution of dust in a
  multi-phase ISM promotes the escape of Ly$\alpha$ photons over
  that of UV continuum photons. It also implies that some combination
  of ISM geometry and kinematics reduces the amount of extinction seen
  by Ly$\alpha$ photons from that expected in a static and homogeneous
  ISM, but only up to the point where it is similar to that
  experienced by continuum photons.

\item The escape fraction of Ly$\alpha$ photons from LAEs, given by
  the ratio between the observed Ly$\alpha$ luminosity and that
  predicted from the dust-corrected UV $SFR$, shows a median value of
  29\%. A large scatter is seen around this number, with objects
  ranging from a few percent to 100\%. Both the median value, and the
  range of observed escape fractions in LAEs, show no evolution across the
  $2<z<4$ range sampled by our objects, and does not seem to evolve
  all the way down to $z=0.3$. Since
  LAE selection is biased to include objects of high escape
  fractions at any combination of dust content, redshift and survey limiting
  luminosity, it is not surprising that this parameter shows little or
  no evolution. This lack of evolution in LAEs does not imply that the
  Ly$\alpha$ escape fraction for the overall galaxy population is not
  evolving.

\item The Ly$\alpha$ escape fraction of LAEs shows a clear correlation with
  $E(B-V)$. The correlation follows what is expected for a value of
  $q=1$, where $q$ is the ratio between the dust opacity seen by
  Ly$\alpha$ and that seen by continuum photons. This behavior is
  consistent with what is observed for LAEs at $z=0.3$, implying that
  not only the value of the escape fraction, but also its dependence
  with dust content, do not evolve with redshift. While other galaxies
  not identified by the LAE selection can have $q>1$, and show low
  Ly$\alpha$ $EW$s, lack of Ly$\alpha$ emission, and even Ly$\alpha$
  in absorption, the lack of objects with $q\ll 1$
  confirms that preferential escape of Ly$\alpha$ photons over
  continuum photons in the presence of a clumpy dust distribution is
  not a common process in galaxies.

\end{itemize} 

We also measure the Ly$\alpha$ luminosity function across our redshift
range. Integration of the luminosity function yields a measurement of
the Ly$\alpha$ luminosity density in our sampled volume. Using our
data, and a compilation of measurements of the Ly$\alpha$ luminosity
function at different redshifts from the literature, we are able
to trace the evolution of $\rho_{Ly\alpha}$ with redshift from
$z=0.3$ to $z=7.7$. Comparing the observed value of $\rho_{Ly\alpha}$
with the expected density implied by the $SFR$ history of the universe, allows a measurement of the
evolution of the average Ly$\alpha$ escape fraction for the overall
galaxy population in this redshift range. From these measurements we
conclude the following:

\begin{itemize}

\item The observed Ly$\alpha$ luminosity function is well matched to
  previous measurements in the literature, especially in the
  $L(Ly\alpha)\leq 10^{43}$ erg s$^{-1}$ range typically
  measured by previous studies. Given the large volume sampled by the
  HETDEX Pilot Survey, we are able to find many objects above this
  luminosity. Both the redshift distribution and the luminosity
  function show hints of a decrease in the number density of LAEs of
  roughly a factor of 2 from $z=4$ to $2$, although this decrease is not
  statistically significant and larger samples are required before it
  can be confirmed. In any case, this decrease goes in the right
  direction and is consistent with what is expected from the observed
  drop in the Ly$\alpha$ escape fraction for the overall galaxy
  population.

\item The Ly$\alpha$ luminosity density is observed to increase
  steadily from $z=0.3$ to $z\simeq 2$, following the behavior of the
  $SFR$ history of the universe. Over this range, the average
  Ly$\alpha$ escape fraction increases very slowly from $\sim1$\%
  to $\sim5$\%. At $z\gtrsim 2$ the increase in $\rho_{Ly\alpha}$ starts
  to flatten, and a decline is observed around $z\sim6$. This
  behavior is accompanied by a decrease in $\rho_{SFR}$ immediately
  after $z=2$, implying that over the $2<z<6$ range, the average
  Ly$\alpha$ escape fraction in galaxies increases steadily from the
  $\sim 5$\%  up to $\sim80$\% by
  $z=6$. Current measurements of the luminosity function at higher
  redshifts do not constrain the behavior of the escape fraction
  beyond $z=6.6$. This drop in the average escape fraction of Ly$\alpha$
  photons with cosmic time is consistent with the increase in the
  dust-content of star forming galaxies, which is expected from the
  chemical enrichment of these objects as star formation proceeds, and
  is observed as a reddening in the UV slope of star forming galaxies
  towards lower redshifts \citep{bouwens10a, finkelstein10}

\item Equation 8 provides a useful analytical form which describes the
  history of the average Ly$\alpha$ escape fraction of galaxies as a
  function of redshift. This equation can prove useful to predict the
  expected average Ly$\alpha$ luminosity of galaxies in
  numerical simulations and semi-analytical models. The reader must
  keep in mind that galaxies do not show a single value of the escape
  fraction at any given redshift, but rather a relatively broad (and
  mostly unconstrained) distribution, so this equation can only be
  used in a statistical sense. It must also be kept in mind that the
  behavior of the escape fraction at $z>6.6$ is still unconstrained.

\end{itemize} 

These last few points have interesting consequences regarding the potential
that observations of LAEs at very high redshifts ($z\geq7$) have to
detect the effects of cosmic re-ionization. Our results imply that at
these redshifts, dust is no longer an important factor setting the
average escape fraction of Ly$\alpha$ photons in galaxies. Therefore,
a significant drop in the Ly$\alpha$ escape fraction could be
more easily interpreted as being caused by an increased neutral
fraction in the IGM.

\begin{figure}[t]
\begin{center}
\epsscale{0.5}
\plotone{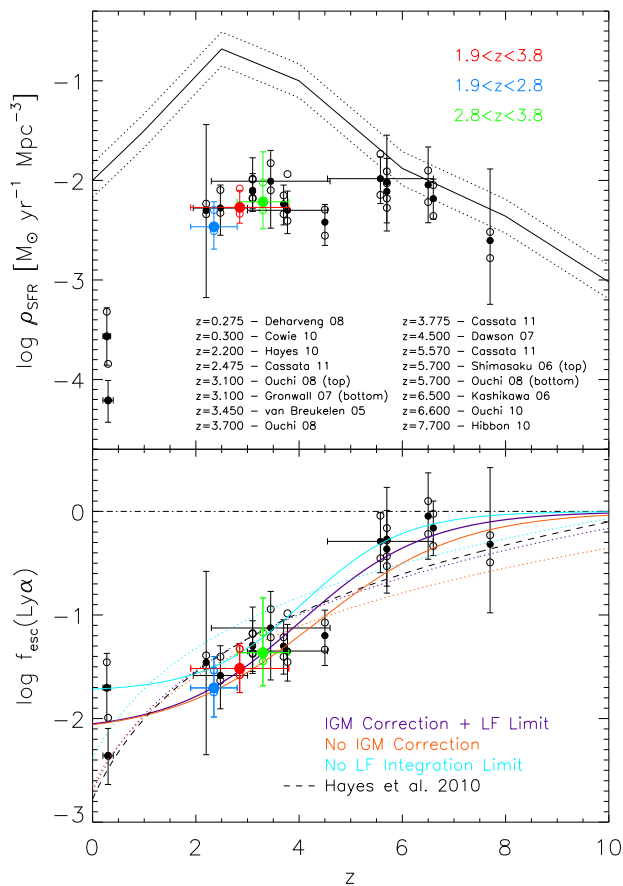}
\caption{{\it Top panel:} $SFR$ density ($\rho_{SFR}$) as a function of redshift. The
  solid and dotted lines show the total $\rho_{SFR}$ from
  \cite{bouwens10b} and its typical uncertainty of 0.17 dex. Blue,
  green, and red filled circles show
  $\rho_{SFR,Ly\alpha}$ derived from the Ly$\alpha$ luminosity function in the
  two redshift bins at $1.9<z<2.8$ and
  $2.8<z<3.8$, as well as for the full sample. Black filled circles show the derived densities at different
  redshifts from the luminosity functions of \cite{vanbreukelen05,
  shimasaku06, kashikawa06, gronwall07, dawson07, 
  ouchi08, deharveng08, ouchi10, cowie10, hayes10a, hibon10}, and
  \cite{cassata11}. Raw values computed without applying an IGM
  correction are shown by the open circles below each
  measurement. Values computed integrating the Ly$\alpha$ luminosity
  functions all the way down to $L(Ly\alpha)=0$ are shown by the open
  circles above each measurement. {\it Bottom panel:} Escape fraction of Ly$\alpha$ photons for the overall galaxy
  population, derived from the ratio between the Ly$\alpha$ derived
  $\rho_{SFR,Ly\alpha}$ and the total value at each redshift. The dashed line
  marks an escape fraction of 100\%. Solid lines shows our best fit
  to the data given by Equation 6, while dotted lines show the best
  fit powerlaw functions. Purple, orange, and cyan colors indicate
  fits to the escape fraction measurements including an IGM
  correction and an integration limit for the luminosity function,
  ignoring the IGM correction, and ignoring the luminosity function
  integration limit respectively. The black dashed line shows the result of \cite{hayes10b}.}
\label{fig-14}
\end{center}
\end{figure}

We thank David Doss, Brian Roman, Kevin Meyer, John Kuehne, and all the
staff at McDonald Observatory for their
invaluable help during the observations. The UV measured SFR history
of the universe shown in Figure \ref{fig-14} was kindly provided by
Richard Bouwens. We thank Hakim Atek for providing his escape fraction
measurements for the $z=0.3$ LAEs. The construction of VIRUS-P was 
possible thanks to the generous support of the Cynthia \& George
Mitchell Foundation. Salary and travel support for students was provided by the Texas Norman
Hackerman Advanced Research Program under grants ARP 003658-0005-2006 and
003658-0295-2007. Guillermo A. Blanc thanks the financial support of Sigma 
Xi, The Scientific Research Society. We also thank Mathew Hayes,
Christian Tapken, Alice Shapley, and the anonymous referee of this
publication, for useful comments and discussions which
helped improve the quality of this work. This research has made use of
NASA's Astrophysics
Data System, and of the NASA/IPAC Extragalactic Database (NED) which
is operated by the Jet Propulsion Laboratory, California Institute of
Technology, under contract with the National Aeronautics and Space
Administration.

\epsscale{1.0}

\clearpage

\input{tab1.tex}

\input{tab2.tex}

\input{tab3.tex}

\end{document}

%% file: tab1.tex
\begin{deluxetable}{cccccccc}
\tabletypesize{\scriptsize}
\tablecaption{Properties of HETDEX Pilot Survey LAEs\label{tbl-1}}
\tablewidth{0pt}
\tablehead{
\colhead{ID\tablenotemark{a}} &  
\colhead{$z$} &
\colhead{$L(Ly\alpha$)} & 
\colhead{$L_{\nu, 1500\AA}$\tablenotemark{b}} & 
\colhead{$\beta$} &
\colhead{$E(B-V)$} &
\colhead{$f_{esc}(Ly\alpha)$} &
\colhead{$EW_0(Ly\alpha)$}}
\startdata
 &  
 &
$10^{42}$erg s$^{-1}$ &
$10^{28}$ erg s$^{-1}$ Hz$^{-1}$ & 
 &
mag &
 &
 \AA\ \\
\tableline\\
   HPS-3   &       3.09   &    14.4$\pm$2.8   &    12.1$\pm$1.6   &    -0.9$\pm$0.4   &    0.27$\pm$0.08   &    0.06$^{+0.08}_{-0.04}$   &    73$\pm$16   \\   
   HPS-6   &       2.78   &    20.1$\pm$2.2   &    19.5$\pm$1.6   &    -1.4$\pm$0.2   &    0.18$\pm$0.06   &    0.12$^{+0.10}_{-0.06}$   &    58$\pm$8   \\   
   HPS-11   &       2.78   &    11.5$\pm$2.5   &    18.7$\pm$1.1   &    -2.2$\pm$0.2   &    0.00$\pm$0.05   &    0.40$^{+0.28}_{-0.09}$   &    28$\pm$6   \\   
   HPS-13   &       3.32   &    10.1$\pm$2.0   &    9.3$\pm$1.9   &    -1.2$\pm$0.5   &    0.20$\pm$0.11   &    0.10$^{+0.17}_{-0.07}$   &    62$\pm$16   \\   
   HPS-17   &       2.78   &    6.9$\pm$1.9   &    13.2$\pm$1.5   &    -2.5$\pm$0.4   &    0.00$\pm$0.08   &    0.34$^{+0.43}_{-0.10}$   &    23$\pm$6   \\   
   HPS-22   &       2.77   &    9.8$\pm$2.7   &    1.9$\pm$1.0   &    -0.6$\pm$1.2   &    0.33$\pm$0.25   &    0.14$^{+1.39}_{-0.14}$   &    340$\pm$187   \\   
   HPS-25   &       2.55   &    32.1$\pm$4.7   &    9.1$\pm$1.2   &    -0.2$\pm$0.3   &    0.41$\pm$0.08   &    0.05$^{+0.05}_{-0.03}$   &    252$\pm$50   \\   
   HPS-34   &       2.76   &    11.2$\pm$2.8   &    11.7$\pm$1.0   &    -1.9$\pm$0.2   &    0.07$\pm$0.06   &    0.33$^{+0.25}_{-0.16}$   &    47$\pm$12   \\   
   HPS-51   &       3.10   &    5.9$\pm$2.9   &    22.4$\pm$2.6   &    -1.0$\pm$0.3   &    0.26$\pm$0.08   &    0.01$^{+0.02}_{-0.01}$   &    16$\pm$8   \\   
   HPS-53   &       3.57   &    13.0$\pm$4.2   &  -   &   -  &   -   &   -   &   -  \\   
   HPS-62   &       2.08   &    17.1$\pm$5.5   &    6.9$\pm$0.8   &    -1.3$\pm$0.3   &    0.18$\pm$0.07   &    0.29$^{+0.29}_{-0.17}$   &    139$\pm$49   \\   
   HPS-82   &       2.25   &    29.1$\pm$7.7   &    1.4$\pm$0.7   &    1.7$\pm$1.4   &    0.79$\pm$0.29   &    0.01$^{+0.11}_{-0.01}$   &    2213$\pm$1548   \\   
   HPS-84   &       3.25   &    24.3$\pm$6.1   &    15.1$\pm$3.8   &    -1.3$\pm$0.5   &    0.19$\pm$0.11   &    0.17$^{+0.33}_{-0.12}$   &    91$\pm$30   \\   
   HPS-89   &       2.54   &    14.4$\pm$3.0   &  -   &   -  &   -   &   -   &   -  \\   
   HPS-91   &       3.00   &    10.3$\pm$3.4   &    23.7$\pm$3.2   &    -0.8$\pm$0.3   &    0.28$\pm$0.08   &    0.02$^{+0.02}_{-0.01}$   &    27$\pm$9   \\   
   HPS-92   &       3.67   &    13.6$\pm$4.4   &    44.9$\pm$5.7   &    -1.6$\pm$0.3   &    0.14$\pm$0.07   &    0.05$^{+0.06}_{-0.03}$   &    16$\pm$5   \\   
   HPS-93   &       2.26   &    20.9$\pm$5.0   &  -   &   -  &   -   &   -   &   -  \\   
   HPS-95   &       2.45   &    13.4$\pm$4.8   &    2.1$\pm$0.6   &    -1.8$\pm$1.2   &    0.09$\pm$0.25   &    1.72$^{+17.09}_{-1.18}$   &    322$\pm$155   \\   
   HPS-99   &       3.01   &    22.8$\pm$4.7   &    6.0$\pm$2.2   &    -0.4$\pm$0.7   &    0.37$\pm$0.16   &    0.07$^{+0.25}_{-0.06}$   &    258$\pm$102   \\   
   HPS-109   &       3.21   &    22.5$\pm$5.9   &    47.5$\pm$5.6   &    -0.9$\pm$0.2   &    0.28$\pm$0.06   &    0.02$^{+0.02}_{-0.01}$   &    29$\pm$8   \\   
   HPS-111   &       3.18   &    11.2$\pm$3.7   &    24.0$\pm$2.9   &    -1.7$\pm$0.3   &    0.11$\pm$0.07   &    0.10$^{+0.10}_{-0.06}$   &    24$\pm$8   \\   
   HPS-124   &       3.74   &    13.0$\pm$6.3   &    10.1$\pm$3.7   &    -3.0$\pm$0.9   &    0.00$\pm$0.19   &    0.83$^{+4.39}_{-0.51}$   &    51$\pm$27   \\   
   HPS-126   &       2.83   &    106.7$\pm$9.1   &    3.5$\pm$3.2   &    1.9$\pm$1.4   &    0.82$\pm$0.28   &    0.01$^{+0.10}_{-0.01}$   &    3338$\pm$3038   \\   
   HPS-127   &       2.54   &    9.0$\pm$3.6   &    10.2$\pm$0.8   &    -1.5$\pm$0.3   &    0.15$\pm$0.08   &    0.14$^{+0.16}_{-0.09}$   &    48$\pm$19   \\   
   HPS-142   &       2.58   &    9.1$\pm$2.2   &    20.2$\pm$1.1   &    -1.1$\pm$0.2   &    0.22$\pm$0.05   &    0.04$^{+0.03}_{-0.02}$   &    26$\pm$6   \\   
   HPS-144   &       2.73   &    2.7$\pm$1.3   &    1.0$\pm$0.5   &    1.3$\pm$1.0   &    0.70$\pm$0.20   &    0.00$^{+0.01}_{-0.00}$   &    270$\pm$187   \\   
   HPS-145   &       2.18   &    26.5$\pm$3.6   &    5.4$\pm$0.5   &    0.1$\pm$0.3   &    0.48$\pm$0.07   &    0.03$^{+0.03}_{-0.02}$   &    380$\pm$66   \\   
   HPS-150   &       2.90   &    18.1$\pm$4.2   &    17.7$\pm$1.2   &    -1.5$\pm$0.2   &    0.15$\pm$0.05   &    0.16$^{+0.11}_{-0.07}$   &    55$\pm$13   \\   
   HPS-153   &       2.71   &    16.3$\pm$3.2   &    5.1$\pm$1.0   &    -0.9$\pm$0.4   &    0.26$\pm$0.08   &    0.18$^{+0.22}_{-0.10}$   &    198$\pm$50   \\   
   HPS-154   &       2.87   &    6.2$\pm$1.6   &    2.5$\pm$1.0   &    -1.1$\pm$0.9   &    0.22$\pm$0.19   &    0.19$^{+1.03}_{-0.17}$   &    148$\pm$64   \\   
   HPS-160   &       2.43   &    6.6$\pm$3.2   &    0.4$\pm$0.4   &    0.1$\pm$2.2   &    0.46$\pm$0.44   &    0.14$^{+9.03}_{-0.15}$   &    1306$\pm$1732   \\   
   HPS-161   &       3.25   &    35.1$\pm$3.7   &    31.6$\pm$2.4   &    -0.4$\pm$0.2   &    0.37$\pm$0.05   &    0.02$^{+0.01}_{-0.01}$   &    76$\pm$9   \\   
   HPS-164   &       2.45   &    10.1$\pm$5.3   &    7.4$\pm$0.8   &    -1.3$\pm$0.2   &    0.20$\pm$0.06   &    0.14$^{+0.13}_{-0.09}$   &    79$\pm$42   \\   
   HPS-168   &       3.45   &    36.4$\pm$3.0   &    7.5$\pm$1.1   &    -2.0$\pm$0.3   &    0.04$\pm$0.07   &    2.08$^{+2.10}_{-0.76}$   &    238$\pm$35   \\   
   HPS-174   &       3.45   &    2.7$\pm$2.0   &    2.1$\pm$0.9   &    -2.5$\pm$1.1   &    0.00$\pm$0.22   &    0.85$^{+5.94}_{-0.73}$   &    58$\pm$45   \\   
   HPS-182   &       2.43   &    10.4$\pm$2.2   &    4.5$\pm$0.4   &    -2.0$\pm$0.3   &    0.04$\pm$0.08   &    0.99$^{+1.11}_{-0.40}$   &    114$\pm$27   \\   
   HPS-183   &       2.16   &    8.6$\pm$5.3   &    3.2$\pm$0.4   &    -2.1$\pm$0.4   &    0.03$\pm$0.09   &    1.36$^{+1.98}_{-0.90}$   &    128$\pm$82   \\   
   HPS-184   &       3.21   &    4.4$\pm$3.0   &    4.4$\pm$1.5   &    -1.8$\pm$0.9   &    0.08$\pm$0.19   &    0.29$^{+1.53}_{-0.26}$   &    51$\pm$36   \\   
   HPS-189   &       2.45   &    4.9$\pm$2.9   &    4.5$\pm$0.6   &    -2.0$\pm$0.3   &    0.05$\pm$0.07   &    0.43$^{+0.50}_{-0.31}$   &    54$\pm$32   \\   
   HPS-190   &       2.28   &    6.0$\pm$1.6   &  -   &   -  &   -   &   -   &   -  \\   
   HPS-194   &       2.29   &    23.5$\pm$1.8   &    10.6$\pm$0.8   &    -1.8$\pm$0.2   &    0.09$\pm$0.06   &    0.62$^{+0.44}_{-0.26}$   &    114$\pm$13   \\   
   HPS-196   &       2.65   &    12.3$\pm$2.0   &    7.4$\pm$1.0   &    0.4$\pm$0.3   &    0.53$\pm$0.07   &    0.01$^{+0.01}_{-0.00}$   &    134$\pm$27   \\   
   HPS-197   &       2.44   &    7.1$\pm$2.6   &    2.2$\pm$1.0   &    -1.9$\pm$1.8   &    0.07$\pm$0.36   &    1.09$^{+32.85}_{-0.70}$   &    160$\pm$93   \\   
   HPS-205   &       2.91   &    12.7$\pm$3.5   &    2.1$\pm$0.9   &    -1.0$\pm$0.9   &    0.26$\pm$0.19   &    0.34$^{+1.71}_{-0.30}$   &    372$\pm$174   \\   
   HPS-207   &       2.71   &    5.0$\pm$1.7   &    2.1$\pm$0.8   &    -2.9$\pm$1.2   &    0.00$\pm$0.25   &    1.54$^{+15.26}_{-0.80}$   &    97$\pm$46   \\   
   HPS-210   &       3.49   &    9.5$\pm$3.0   &    9.6$\pm$2.4   &    -1.3$\pm$0.5   &    0.18$\pm$0.11   &    0.11$^{+0.19}_{-0.08}$   &    56$\pm$20   \\   
   HPS-213   &       3.30   &    11.0$\pm$2.8   &    11.6$\pm$1.3   &    -0.5$\pm$0.3   &    0.35$\pm$0.07   &    0.02$^{+0.02}_{-0.01}$   &    63$\pm$17   \\   
   HPS-214   &       3.30   &    6.6$\pm$3.1   &    1.4$\pm$0.5   &    -2.7$\pm$1.2   &    0.00$\pm$0.24   &    3.11$^{+28.94}_{-1.77}$   &    202$\pm$102   \\   
   HPS-223   &       2.31   &    12.9$\pm$3.5   &    2.0$\pm$0.5   &    -1.2$\pm$0.6   &    0.21$\pm$0.13   &    0.55$^{+1.41}_{-0.42}$   &    373$\pm$142   \\   
   HPS-229   &       3.04   &    31.6$\pm$3.5   &    30.5$\pm$1.9   &    -1.6$\pm$0.2   &    0.14$\pm$0.05   &    0.18$^{+0.12}_{-0.07}$   &    55$\pm$6   \\   
   HPS-231   &       2.72   &    16.1$\pm$4.1   &    1.8$\pm$0.5   &    -1.9$\pm$0.8   &    0.07$\pm$0.18   &    2.95$^{+12.73}_{-1.71}$   &    459$\pm$190   \\   
   HPS-244   &       2.10   &    2.6$\pm$1.2   &    1.7$\pm$0.4   &    -2.3$\pm$0.7   &    0.00$\pm$0.15   &    1.00$^{+3.04}_{-0.50}$   &    71$\pm$38   \\   
   HPS-249   &       3.27   &    5.7$\pm$2.2   &    2.5$\pm$0.8   &    -2.6$\pm$0.7   &    0.00$\pm$0.14   &    1.48$^{+4.30}_{-0.75}$   &    98$\pm$44   \\   
   HPS-251   &       2.29   &    14.3$\pm$4.0   &    5.2$\pm$0.5   &    -1.9$\pm$0.3   &    0.07$\pm$0.08   &    0.88$^{+0.95}_{-0.51}$   &    140$\pm$43   \\   
   HPS-253   &       3.18   &    15.4$\pm$3.0   &    12.9$\pm$1.4   &    -1.7$\pm$0.2   &    0.10$\pm$0.06   &    0.29$^{+0.24}_{-0.14}$   &    62$\pm$13   \\   
   HPS-256   &       2.49   &    13.9$\pm$3.5   &    3.5$\pm$0.6   &    -1.7$\pm$0.4   &    0.10$\pm$0.09   &    1.01$^{+1.49}_{-0.65}$   &    206$\pm$65   \\   
   HPS-258   &       2.81   &    19.3$\pm$2.4   &    13.8$\pm$0.9   &    -0.8$\pm$0.2   &    0.28$\pm$0.06   &    0.06$^{+0.05}_{-0.03}$   &    88$\pm$12   \\   
   HPS-263   &       2.43   &    9.2$\pm$3.0   &    9.7$\pm$0.6   &    -1.7$\pm$0.2   &    0.10$\pm$0.06   &    0.23$^{+0.18}_{-0.12}$   &    49$\pm$16   \\   
   HPS-266   &       2.20   &    13.8$\pm$1.9   &  -   &   -  &   -   &   -   &   -  \\   
   HPS-269   &       2.57   &    6.2$\pm$1.6   &    3.6$\pm$0.5   &    -1.9$\pm$0.5   &    0.07$\pm$0.10   &    0.57$^{+0.98}_{-0.32}$   &    87$\pm$26   \\   
   HPS-273   &       3.64   &    15.0$\pm$5.7   &  -   &   -  &   -   &   -   &   -  \\   
   HPS-274   &       2.87   &    10.7$\pm$2.0   &    9.8$\pm$0.9   &    -1.3$\pm$0.2   &    0.19$\pm$0.06   &    0.11$^{+0.08}_{-0.05}$   &    62$\pm$12   \\   
   HPS-283   &       3.30   &    19.3$\pm$2.8   &    14.4$\pm$1.6   &    -1.3$\pm$0.2   &    0.20$\pm$0.06   &    0.14$^{+0.11}_{-0.06}$   &    76$\pm$13   \\   
   HPS-286   &       2.23   &    9.2$\pm$2.0   &    7.6$\pm$1.5   &    -2.1$\pm$0.5   &    0.03$\pm$0.11   &    0.61$^{+1.13}_{-0.21}$   &    59$\pm$18   \\   
   HPS-287   &       3.32   &    4.7$\pm$2.1   &    2.4$\pm$0.9   &    -1.5$\pm$1.2   &    0.14$\pm$0.24   &    0.33$^{+2.88}_{-0.29}$   &    107$\pm$55   \\   
   HPS-288   &       3.04   &    8.4$\pm$2.1   &    12.0$\pm$1.2   &    -1.8$\pm$0.3   &    0.10$\pm$0.06   &    0.18$^{+0.16}_{-0.09}$   &    36$\pm$9   \\   
   HPS-292   &       2.87   &    19.6$\pm$2.6   &    6.3$\pm$0.8   &    -1.6$\pm$0.4   &    0.12$\pm$0.08   &    0.63$^{+0.75}_{-0.35}$   &    166$\pm$30   \\   
   HPS-296   &       2.84   &    5.8$\pm$2.2   &    6.3$\pm$1.0   &    -1.6$\pm$0.4   &    0.13$\pm$0.10   &    0.17$^{+0.28}_{-0.12}$   &    49$\pm$19   \\   
   HPS-306   &       2.44   &    14.8$\pm$2.9   &    11.8$\pm$0.8   &    -1.9$\pm$0.2   &    0.07$\pm$0.05   &    0.43$^{+0.29}_{-0.19}$   &    62$\pm$13   \\   
   HPS-310   &       3.07   &    7.6$\pm$1.9   &    7.2$\pm$1.2   &    -1.5$\pm$0.4   &    0.15$\pm$0.09   &    0.16$^{+0.21}_{-0.10}$   &    58$\pm$16   \\   
   HPS-313   &       2.10   &    6.7$\pm$3.0   &    24.9$\pm$0.8   &    -1.5$\pm$0.1   &    0.15$\pm$0.04   &    0.04$^{+0.03}_{-0.02}$   &    14$\pm$6   \\   
   HPS-314   &       2.63   &    6.9$\pm$2.1   &  -   &   -  &   -   &   -   &   -  \\   
   HPS-315   &       3.07   &    5.9$\pm$1.8   &    14.7$\pm$1.5   &    -1.5$\pm$0.3   &    0.14$\pm$0.07   &    0.07$^{+0.06}_{-0.04}$   &    21$\pm$6   \\   
   HPS-316   &       2.81   &    13.1$\pm$3.4   &    13.2$\pm$1.1   &    -2.0$\pm$0.2   &    0.04$\pm$0.06   &    0.44$^{+0.36}_{-0.18}$   &    48$\pm$13   \\   
   HPS-318   &       2.46   &    11.6$\pm$3.8   &    13.4$\pm$0.8   &    -1.3$\pm$0.1   &    0.18$\pm$0.05   &    0.10$^{+0.07}_{-0.05}$   &    49$\pm$16   \\   
   HPS-327   &       2.25   &    4.7$\pm$1.9   &  -   &   -  &   -   &   -   &   -  \\   
   HPS-338   &       2.60   &    15.2$\pm$4.0   &    1.7$\pm$0.9   &    -1.9$\pm$1.1   &    0.06$\pm$0.23   &    3.30$^{+25.40}_{-1.94}$   &    452$\pm$260   \\   
   HPS-341   &       2.93   &    8.4$\pm$2.3   &    8.0$\pm$1.2   &    -2.2$\pm$0.5   &    0.01$\pm$0.10   &    0.61$^{+1.00}_{-0.20}$   &    50$\pm$15   \\   
   HPS-360   &       2.92   &    11.5$\pm$3.0   &    7.1$\pm$2.0   &    -1.3$\pm$0.5   &    0.19$\pm$0.12   &    0.18$^{+0.37}_{-0.13}$   &    91$\pm$33   \\   
   HPS-370   &       3.18   &    8.7$\pm$2.5   &    5.2$\pm$1.3   &    -2.0$\pm$0.6   &    0.04$\pm$0.12   &    0.75$^{+1.61}_{-0.34}$   &    81$\pm$28   \\   
   HPS-372   &       2.76   &    5.5$\pm$1.4   &    1.4$\pm$1.3   &    -1.9$\pm$2.5   &    0.07$\pm$0.51   &    1.32$^{+165.25}_{-0.94}$   &    194$\pm$183   \\   
   HPS-373   &       2.91   &    11.3$\pm$2.6   &  -   &   -  &   -   &   -   &   -  \\   
   HPS-389   &       2.59   &    10.2$\pm$1.9   &    7.9$\pm$1.1   &    -1.5$\pm$0.3   &    0.14$\pm$0.08   &    0.21$^{+0.25}_{-0.12}$   &    70$\pm$16   \\   
   HPS-391   &       2.96   &    17.4$\pm$4.1   &    8.4$\pm$2.0   &    -1.6$\pm$0.5   &    0.12$\pm$0.10   &    0.44$^{+0.72}_{-0.29}$   &    110$\pm$34   \\   
   HPS-395   &       2.27   &    6.6$\pm$2.8   &    10.2$\pm$1.2   &    -1.9$\pm$0.3   &    0.07$\pm$0.07   &    0.22$^{+0.22}_{-0.14}$   &    32$\pm$14   \\   
   HPS-402   &       2.97   &    11.2$\pm$1.6   &    5.1$\pm$1.2   &    -2.1$\pm$0.6   &    0.02$\pm$0.13   &    1.20$^{+2.78}_{-0.36}$   &    105$\pm$28   \\   
   HPS-403   &       3.18   &    7.5$\pm$1.6   &    9.6$\pm$1.9   &    -1.9$\pm$0.4   &    0.08$\pm$0.10   &    0.25$^{+0.38}_{-0.14}$   &    39$\pm$10   \\   
   HPS-415   &       3.37   &    10.5$\pm$2.5   &    6.1$\pm$1.2   &    -2.0$\pm$0.5   &    0.06$\pm$0.11   &    0.65$^{+1.19}_{-0.32}$   &    86$\pm$25   \\   
   HPS-419   &       2.24   &    8.1$\pm$1.4   &    6.0$\pm$0.8   &    -1.7$\pm$0.4   &    0.12$\pm$0.08   &    0.29$^{+0.36}_{-0.17}$   &    71$\pm$15   \\   
   HPS-420   &       2.93   &    12.1$\pm$2.5   &    5.5$\pm$1.7   &    -1.3$\pm$0.6   &    0.19$\pm$0.13   &    0.22$^{+0.52}_{-0.16}$   &    125$\pm$47   \\   
   HPS-426   &       3.41   &    6.6$\pm$1.6   &    6.4$\pm$1.5   &    -1.3$\pm$0.5   &    0.18$\pm$0.10   &    0.12$^{+0.19}_{-0.08}$   &    58$\pm$18   \\   
   HPS-428   &       3.34   &    13.0$\pm$2.3   &    22.0$\pm$2.3   &    -1.4$\pm$0.3   &    0.16$\pm$0.07   &    0.08$^{+0.08}_{-0.04}$   &    32$\pm$16   \\   
   HPS-434   &       2.27   &    3.9$\pm$1.2   &    1.0$\pm$0.4   &    -2.5$\pm$1.6   &    0.00$\pm$0.32   &    2.65$^{+51.32}_{-1.36}$   &    180$\pm$139   \\   
   HPS-436   &       2.42   &    2.7$\pm$1.0   &    4.1$\pm$0.7   &    -2.8$\pm$0.6   &    0.00$\pm$0.12   &    0.42$^{+0.94}_{-0.16}$   &    27$\pm$10   \\   
   HPS-447   &       3.13   &    5.0$\pm$1.1   &    14.8$\pm$2.1   &    -1.6$\pm$0.3   &    0.13$\pm$0.08   &    0.06$^{+0.07}_{-0.04}$   &    17$\pm$4   \\   
   HPS-462   &       2.21   &    27.4$\pm$2.9   &    8.3$\pm$0.8   &    -1.8$\pm$0.3   &    0.08$\pm$0.08   &    0.98$^{+1.10}_{-0.53}$   &    169$\pm$27   \\   
   HPS-466   &       3.24   &    18.2$\pm$2.1   &    31.3$\pm$3.4   &    -1.5$\pm$0.2   &    0.15$\pm$0.06   &    0.09$^{+0.07}_{-0.04}$   &    32$\pm$4   \\   
   HPS-467   &       2.80   &    5.0$\pm$1.8   &  -   &   -  &   -   &   -   &   -  \\   
   HPS-474   &       2.28   &    4.3$\pm$2.4   &    3.9$\pm$0.4   &    -1.9$\pm$0.3   &    0.07$\pm$0.07   &    0.36$^{+0.39}_{-0.27}$   &    56$\pm$32   \\   
\enddata
\tablenotetext{a}{ID corresponds to that in Table 3 of \cite{adams11}. Equatorial coordinates and line fluxes are provided there.}
\tablenotetext{b}{Dashes indicate objects with no broad-band counterpart.}
\end{deluxetable}

%% file: tab2.tex
\begin{deluxetable}{cccccc}
\tabletypesize{\scriptsize}
\tablecaption{Ly$\alpha$ luminosity function Best Fit Schechter Parameters, Luminosity and $SFR$ Density\label{tbl-2}}
\tablewidth{0pt}
\tablehead{
\colhead{Sample} & \colhead{$\alpha$\tablenotemark{a}} & \colhead{$\phi^*$} &
\colhead{L$^*$} & \colhead{$\rho_{Ly\alpha}$} & \colhead{$\rho_{SFR,Ly\alpha}$}
}
\startdata
 & &  $10^{-4}$Mpc$^{-3}$ & $10^{43}$erg s$^{-1}$ & $10^{39}$erg s$^{-1}$Mpc$^{-3}$ & $10^{-3}$M$_{\odot}$yr$^{-1}$Mpc$^{-3}$\\
\tableline\\
$1.9<z<3.8$                           & $-1.7$ & $2.2_{-1.3}^{+3.9}$  & $1.20_{-0.52}^{+1.02}$ & $5.1_{-1.6}^{+2.5}$   & $4.6_{-1.4}^{+2.2}$ \\ 
$1.9<z<3.8$                           & $-1.5$ & $2.9_{-1.7}^{+4.4}$  & $1.01_{-0.41}^{+0.67}$ & $4.3_{-1.3}^{+2.0}$   & $3.9_{-1.2}^{+1.8}$ \\  
$1.9<z<3.8$, $L(Ly\alpha)\leq10^{43}$ & $-1.7$ & $6.7_{-5.9}^{+30.6}$ & $0.60_{-0.33}^{+2.99}$ & $6.8_{-2.7}^{+7.6}$   & $6.2_{-2.5}^{+6.9}$ \\
$1.9<z<2.8$                           & $-1.7$ & $1.0_{-0.9}^{+5.4}$  & $1.63_{-1.08}^{+9.46}$ & $3.4_{-1.4}^{+2.7}$   & $3.1_{-1.3}^{+2.4}$\\  
$2.8\leq z<3.8$                       & $-1.7$ & $2.6_{-2.2}^{+28.3}$ & $1.11_{-0.74}^{+2.40}$ & $5.5_{-2.6}^{+12.0}$  & $5.0_{-2.3}^{+10.9}$\\

\enddata
\tablenotetext{a}{Fixed parameter}
\end{deluxetable}

%% file: tab3.tex
\begin{deluxetable}{ccccccc}
\tabletypesize{\scriptsize}
\tablecaption{Ly$\alpha$ Escape Fraction History Best Fit Paramenters\label{tbl-3}}
\tablewidth{0pt}
\tablehead{
\colhead{Function} & 
\colhead{Data Points} & 
\colhead{${\rm log}(f_{esc}(0))$} &
\colhead{$\xi$} & 
\colhead{$z_{tr}$} &
\colhead{$\theta$} &
\colhead{$\chi_{red}^2$}
}
\startdata
           & IGM corr + LF limit & $-2.7\pm0.2$ & $2.4\pm0.3$ & - & - & $1.1$ \\
Power Law  & No IGM corr         & $-2.7\pm0.2$ & $2.2\pm0.3$ & - & - & $1.0$ \\
           & No LF limit         & $-2.4\pm0.2$ & $2.2\pm0.3$ & - & - & $1.2$ \\
\tableline\\
           & IGM corr + LF limit & $-2.1\pm0.3$ & - & $4.0\pm0.5$ & $0.4\pm0.1$ & $0.41$ \\
Transition & No IGM corr         & $-2.2\pm0.3$ & - & $4.3\pm0.6$ & $0.4\pm0.1$ & $0.38$ \\
           & No LF limit         & $-1.7\pm0.2$ & - & $4.1\pm0.4$ & $0.5\pm0.2$ & $0.39$ \\
\enddata
\end{deluxetable}